\documentclass[aip,jcp,amsmath, amssymb, 
reprint,
twocolumn
]{revtex4-1}
\usepackage{graphicx}
\usepackage{xcolor}
\usepackage[hypertexnames=false]{hyperref}
\hypersetup{
    colorlinks,
    linkcolor={black!50!black},
    citecolor={black!50!black},
    urlcolor={blue!80!black}
}

\begin{document}
\raggedbottom

\title{\emph{Operando} XANES from first-principles and its application to iridium oxide} 

\author{Francesco Nattino}
\email{francesco.nattino@epfl.ch}
\affiliation{Theory and Simulations of Materials (THEOS) and National Centre for Computational Design 
and Discovery of Novel Materials (MARVEL), \'{E}cole Polytechnique F\'{e}d\'{e}rale de Lausanne, 
CH-1015 Lausanne, Switzerland.}

\author{Nicola Marzari}
\affiliation{Theory and Simulations of Materials (THEOS) and National Centre for Computational Design 
and Discovery of Novel Materials (MARVEL), \'{E}cole Polytechnique F\'{e}d\'{e}rale de Lausanne, 
CH-1015 Lausanne, Switzerland.}

\date{\today} 

\begin{abstract}
Efficient electro-catalytic water-splitting technologies require suitable catalysts for the oxygen evolution reaction (OER). The development of novel catalysts could benefit from the achievement of a complete understanding of the reaction mechanism on iridium oxide (IrO$_2$), an active catalyst material that is, however, too scarce for large-scale applications. Considerable insight has already been provided by \emph{operando} X-ray absorption near-edge structure (XANES) experiments, which paved the way towards an atomistic description of the catalyst's evolution in a working environment. We combine here first-principles simulations augmented with a continuum description of the solvent and electrolyte to investigate the electrochemical stability of various IrO$_2$ interfaces and to predict the XANES cross-section for selected terminations under realistic conditions of applied potential. The comparison of computed O K-edge XANES spectra to corresponding experiments supports the formation of electron-deficient surface oxygen species in the OER-relevant voltage regime. Furthermore, surface hydroxyl groups that are found to be stable up to $\sim$1 V are suggested to be progressively oxidized at larger potentials, giving rise to a shift in the Ir L$_3$-edge cross-section that qualitatively agrees with measurements.
\end{abstract}

\maketitle

\section{Introduction\label{sec:introduction}}

Hydrogen production through electrocatalytic water splitting is one of the envisaged strategies for the conversion and storage of energy coming from renewable sources. Various materials are known to efficiently catalyze the hydrogen-evolution reaction, which is the cathodic half-reaction in the water splitting process. In contrast, very few catalysts are known for the anodic sub-process, i.e. the oxygen-evolution reaction (OER)\cite{Fabbri2014}. For this half-reaction, reasonable current densities are typically observed only at large overpotentials. In addition, proton exchange-membranes, which are the state-of-the-art electrolysis setups, require harsh acidic conditions, challenging the stability of many potentially interesting catalysts. Iridium oxide (IrO$_2$) represents one of the few stable catalyst materials, being able to provide a reasonable catalytic activity while withstanding the corrosive environment in which the OER takes place. 

Many studies focused on the characterization of the IrO$_2$ interface under electrochemical conditions, with the goal of obtaining an atomistic understanding of the factors that make this catalyst the current gold standard for the OER. This knowledge is expected to help in designing new catalysts that, ideally, would be based on earth-abundant elements, as desirable for large-scale applications. Considerable insight on the electronic structure of iridium oxide has been provided by X-ray absorption spectroscopy. While few studies have explored the bulk and surface properties of iridium-oxide-based catalysts using \emph{ex-situ} techniques\cite{Kotz1984}, recent developments of electrochemical-cell setups at synchrotron-radiation sources\cite{Binninger2016, Frevel2019} have enabled \emph{operando} investigations in a realistic working environment. Most studies made use of  soft X-rays, which guarantee higher surface sensitivity, thus focusing on the oxygen K-edge in X-ray absorption near-edge spectroscopy (XANES). However, higher-energy X-rays have also been used in combination with high-surface-area IrO$_2$ nanoparticles or thin films, providing direct information on the iridium oxidation state by probing the corresponding L$_2$/L$_3$-edge\cite{Minguzzi2014, Binninger2016, Abbott2016}.

\emph{In-situ} XANES experiments have allowed to identify potentially active sites in IrO$_2$-derived catalysts. Isotopic-labeling experiments have revealed that surface oxygen atoms directly participate to the OER\cite{Fierro2007}. Later on, Pfeifer \emph{et al.} have first reported the presence of electrophilic oxygen (namely O$^{\mathrm{I}-}$) in amorphous IrO$_x$\cite{Pfeifer2016, Pfeifer2016b}. This species, which appears as a pre-edge resonance in O K-edge spectra, has been proposed to be highly reactive towards nucleophilic attack\cite{Pfeifer2016a} and therefore suggested to play a major role in the formation of O-O bonds in the OER. \emph{Operando} XANES experiments have provided additional evidence for this hypothesis, showing that electrophilic oxygen species can be observed during oxygen-evolution both on electrochemically-oxidized iridium nanoparticles\cite{Pfeifer2017, Knop-Gericke2017, Saveleva2018, Frevel2019} and on thermally-synthesized IrO$_2$ samples\cite{Saveleva2018}. 

First-principles simulations have played a determining role in identifying the spectroscopic features that correspond to such reactive oxygen site, enabling the interpretation of corresponding XANES cross-sections. Density-functional theory (DFT) simulations of oxygen-rich bulk\cite{Pfeifer2016, Pfeifer2016b} and surface \cite{Pfeifer2016a, Pfeifer2017, Saveleva2018, Frevel2019} environments have predicted under-coordinated oxygen atoms to show up as resonances at $528-529$ eV in the O K-edge, matching the photon energies at which the pre-edge absorption peaks are observed experimentally. In addition to computational spectroscopies, simulations have also been employed to investigate potential reaction pathways for the OER\cite{Rossmeisl2007, Hansen2010, Ping2015, Siahrostami2015, Ping2017, Gauthier2017}. Most studies considered the (110) facet of rutile IrO$_2$ as a model catalyst surface, motivated by the fact that vacuum DFT calculations predict it to be the minimum-energy termination\cite{Ping2015}. This picture has been recently revised by Opalka \emph{et al.}, who have conducted a thorough study of the stability of low-index IrO$_2$ terminations\cite{Opalka2019}. By using the computational hydrogen electrode approach to account for voltage effects and implicit solvation, they have found the widely-studied (110) surface to be the minimum-energy termination only at low and moderate applied potentials. In the high-potential regime, where oxygen evolution actually takes place, the (111) termination has been found instead to be most stable facet.  

In this work, we combine DFT simulations with a continuum description of solvent and electrolyte to realistically  investigate the interface stability of various IrO$_2$ terminations while simultaneously predicting oxygen K-edge and iridium L$_3$-edge XANES cross-sections in the model electrochemical environment. For the interface stability analysis we exploit a full grand-canonical description of protons and electrons, so that pH and voltage effects are fully decoupled\cite{Hormann2019}. Explicitly-charged interfaces, which realistically represent the oxide surfaces under applied potential conditions, are thus simulated. The continuum description of the electrolyte solution allows to accurately reproduce long-range screening effects while simultaneously avoiding the pitfalls involved in the widely-employed static-solvent descriptions. Finally, the simulation of X-ray absorption cross-sections at finite applied potentials in the implicit electrochemical environment allows us to directly compare simulations to \emph{operando} experiments. To the best of our knowledge, this is the first time that potential-dependent first-principles-based XANES simulations are implemented.  

In agreement with Opalka \emph{et al.}, we find that the (110) and (111) terminations are the most stable interfaces, respectively, at intermediate or high applied potentials\cite{Opalka2019}. We find, however, a reconstructed Ir-rich (101) termination to be the lowest-energy surface at open-circuit conditions. The comparison of simulated XANES cross-sections to experiments at the oxygen K-edge supports the formation of electron-deficient O species on the catalyst surface\cite{Pfeifer2016a, Pfeifer2017, Saveleva2018, Frevel2019}. Furthermore, the shift in Ir L$_3$-edge cross-sections that we predict during the progressive oxidation of the hydroxylated (110) and (101) surfaces agrees with measured trends\cite{Abbott2016}.

The paper is organized as follows. The methodology is presented in Section \ref{sec:methods}, where we briefly describe the approach employed to investigate the electrochemical stability of the various IrO$_2$ terminations (Section \ref{sub:stability}) and the technique used to compute XANES cross-sections (Section \ref{sub:xanes}). All computational details are then provided in Section \ref{sub:computational-details}. Results are presented and discussed in Section \ref{sec:results}, with  XANES simulations on bulk IrO$_2$ presented  in Section \ref{sub:bulk}, results on the electrochemical stability of the various interfaces given in Section \ref{sub:surf-stability} and  \emph{operando} XANES simulations illustrated in Section \ref{sub:XANES-results} (oxygen-K-edge- and iridium-L$_3$-edge-results are presented in Section \ref{subsub:OKedge} and Section \ref{subsub:IrL3edge}, respectively). Finally, Section \ref{sec:conclusions} includes a summary of the paper and the conclusions.

\section{Methods\label{sec:methods}}

\subsection{Electrochemical Stability\label{sub:stability}}

The electrochemical stability of the IrO$_2$ terminations considered has been estimated with the grand-canonical approach described in Ref.\cite{Hormann2019}. Briefly, slab total energies have been first converted to free energies using \emph{ab-initio} thermodynamics approaches\cite{Reuter2001, Reuter2003} and then employed to estimate slab formation energies with respect to \emph{reservoir} components: bulk IrO$_2$, water, protons and electrons. Vibrational contributions to the free energies have been accounted approximately, including zero-point energy corrections for the only hydrogen-containing species, which are expected to be the most significant. Formation energies have been normalized with respect to surface area so that the resulting surface free energies can be straightforwardly compared to each other. The most stable interface at a given applied potential $U$ has been obtained by minimizing the system free energy with respect to interface configuration and surface charge. We have thus modeled explicitly charged terminations by adding (removing) electrons to (from) the systems and determined the corresponding applied potential \emph{a posteriori} as the difference between the asymptotic electrostatic potential (set to zero) and the Fermi energy of the system $\epsilon_F$. Overall, the surface free energy $\gamma$ for a given termination of IrO$_2$ with surface area $A$ can be written as\cite{Hormann2019}:
\begin{equation}
\gamma 	= \frac{1}{2A}\left(\Delta G_{\mathrm{slab}} - n_{\mathrm{H^+_{(aq)}}}\mu_{\mathrm{H^+_{(aq)}}} - n_{\mathrm{e^-}}\mu_{\mathrm{e^-}}\right),
\label{eq:surface-energy}
\end{equation}
where $\Delta G_{\mathrm{slab}}$ is the formation free energy of the slab from bulk IrO$_2$ and water, and $n_{\mathrm{H^+_{(aq)}}}$ and $n_{\mathrm{e^-}}$ are the number of protons and electrons, respectively, which are additionally required to form the (potentially charged) interface. Centro-symmetric slabs that exhibit two identical terminations have been employed, giving rise to the factor two at the denominator in Equation \ref{eq:surface-energy}. The proton and the electron chemical potentials ($\mu_{\mathrm{H^+_{(aq)}}}$ and $\mu_{\mathrm{e^-}}$, respectively), are computed as\cite{Hormann2019}:
\begin{align}
\mu_{\mathrm{H^+_{(aq)}}} &=  \frac{1}{2}\mu_{\mathrm{H}_{2(g)}} + 4.44\ \mathrm{eV} - k_BT\ln 10\ \mathrm{pH},\\
\mu_{\mathrm{e^-}} &= -\epsilon_F = -e \left(U + 4.44\ \mathrm{V}\right),
\end{align}
where $k_B$ is the Boltzmann constant, $T$ is the temperature and $e$ the elementary charge. The applied potential $U$ is expressed here on the standard hydrogen electrode (SHE) scale, whose zero on the absolute scale (i.e. with respect to vacuum) has been estimated to 4.44 V. Thus, exactly as in the well-known computational hydrogen electrode (CHE) approach, the calculation of the chemical potential of the solvated proton is bypassed by exploiting the equilibrium of this species with gaseous H$_2$\cite{Norskov2004}. However, in contrast with the CHE, $n_{\mathrm{e^-}}$ is not bound to $n_{\mathrm{H^+_{(aq)}}}$\cite{Hormann2019}, so that pH and potential effects are fully decoupled. Thus, as noted in Ref.\cite{Hormann2019}, this method  represents a generalization of the CHE approach, which was employed e.g. in Refs. \cite{Matz2017, Pfeifer2017, Opalka2019}. Note that Ping \emph{et al.} have investigated the deprotonation of the (110) IrO$_2$ termination with an analogous grand-canonical treatment of electrons and protons, using, however, a constant-potential framework instead of the constant-charge framework described here\cite{Ping2017}.

Both solvent and electrolyte effects are accounted for, separately, at a continuum level. Explicit water models have been tested on IrO$_2$ interfaces, using either standard optimization techniques\cite{Siahrostami2015, Ping2017} or a more advanced minima-hopping algorithm\cite{Gauthier2017} to identify the lowest energy structures of the water layers at the interface. Results indicate relatively weak solvation effects, with the binding energy of only few hydrogen-donor adsorbates (like OOH) being influenced by the presence of explicit water. These findings validate the approximate treatment of solvation effects with a continuum model. The modeling of the diffuse layer is mandatory here  in order to guarantee the charge neutrality of the system. Explicit electrolyte models would require prohibitively large supercells in order to give access to potential variations in sufficiently small steps. In contrast, the continuum description of the electrochemical environment allows to straightforwardly account for solution-related free-energy contributions that arise from the formation of the diffuse layer. As typically carried out in this class of models, these contributions are included in the density-functional total energy\cite{Borukhov1997, Ringe2016, Nattino2019}. 

\subsection{X-Ray Absorption Near-Edge Structure\label{sub:xanes}}

XANES spectra have been computed using the approach of Refs.\cite{Taillefumier2002, Gougoussis2009a, Bunau2013}, as implemented in the XSpectra package in the \textsc{Quantum ESPRESSO} (QE) distribution\cite{Giannozzi2009, Giannozzi2017}. Briefly, the cross section has been calculated from single-particle states using Fermi's golden rule and the dipole approximation:
\begin{equation}
\sigma(\omega)=4\pi^2\alpha\hbar\omega\sum_f\left|\langle\psi_f|\hat{e}\cdot\textbf{r}|\psi_i\rangle\right|^2\delta(E_f - E_i - \hbar\omega),
\label{eq:XANES}
\end{equation}
where $\alpha$ is the hyperfine constant, $\hat{e}$ is the polarization of the incident radiation and $\textbf{r}$ is the position vector. For the initial state $|\psi_i\rangle$, with orbital energy $E_i$, we have employed the suitable core-state wave-function that we have calculated using the atomic code in QE. The final states $\{|\psi_f\rangle\}$ (with orbital energies $\{E_f\}$) are taken from pseudo-potential-based SCF calculations where we also account for the presence of the continuum solvent and electrolyte, as suitable to model $operando$ electrochemical conditions (see also Section \ref{sub:computational-details}).

The final states have been computed by neglecting the core-hole that follows the absorption of the X-ray radiation. Due to the strong core-hole screening that takes place in metallic systems\cite{Bunau2013}, this approximation has been shown to be very accurate for systems like IrO$_2$\cite{Pfeifer2016, Pfeifer2016a, Pfeifer2016b}. The summation in Equation \ref{eq:XANES} is meant to run only over empty states, i.e. the states above the Fermi level $\epsilon_{F}$ with a smoothing function applied in a narrow interval around this transition ($\pm 10$ meV). The cross-section peaks have been broadened with a frequency-dependent Lorentzian that accounts for the finite lifetime of the initial and final states. The Lorentzian width $\Gamma$ has been assumed to increase linearly with the frequency: $\Gamma = \Gamma_0 + \kappa(\hbar\omega - \epsilon_{F})$. For the oxygen K-edge, we have used the same parameters as in Refs. \cite{Pfeifer2016, Pfeifer2016a, Pfeifer2016b} ($\Gamma_0 = 0.14$ eV, $\kappa = 0.1$). The values of $\Gamma_0$ that we have employed to account for the lifetime of the iridium core-hole in the corresponding L$_3$-edge cross sections have been taken as the width of the experimental white lines of powder samples\cite{Clancy2012}: $\Gamma_0=2.4$ eV for IrO$_2$ and $\Gamma_0=3.3$ eV for Ir. All computed cross sections have been averaged over three orthogonal incident polarization directions, as suitable to compare to experimental investigations that focused either on powders or on heterogeneous samples.
 
The cross sections computed for the various surface terminations, atomic sites and charge states have been translated to an absolute energy scale by using the corresponding core-electron binding energies (BEs)\cite{Pfeifer2016, Pfeifer2016a, Pfeifer2016b, Saveleva2018, Frevel2019}. For each surface structure, charge and symmetry-inequivalent atomic site, the core-electron BE of the absorbing atom has been computed with the $\Delta$SCF method\cite{Pehlke1993, Bianchettin2006}, using core-hole pseudo-potentials to estimate the total energy of the final (core excited) state. Relative BEs have been computed with respect to bulk IrO$_2$, for which we have taken BE values that best align bulk theoretical cross sections to experimental  spectra on powder samples. In particular, we have used the bulk BE values of 530 eV and 11218.5 eV for the oxygen 1s and the iridium $2p_{\frac{3}{2}}$ electrons, respectively. Note that the final-state calculation should be performed in a supercell in order to minimize core-hole interactions and to obtain size-converged absolute BEs. However, only relative BEs are required here, i.e. BE differences between slab and bulk configurations ($\Delta$BEs), for which  size-converged values are obtained even within the primitive cell (see Figure S1 and S2 in the ESI\dag).

The spin-orbit coupling (SOC) splits the iridium core $2p$ levels into $2p_{\frac{1}{2}}$ and $2p_{\frac{3}{2}}$ sub-levels. A framework based on fully-relativistic pseudo-potentials should in principle be used to directly simulate absorption processes from either of these levels. We neglect here SOC effects on the final states and we use the wave-function computed for the isolated Ir atom with scalar relativistic corrections as initial state. These approximations have been shown to be accurate for transition metals of the $3d$ and $4d$ series like Cu and Mo\cite{Bunau2013}. While these approximations could be more severe for $5d$ elements like Ir, we have verified that the neglect of SOC on the initial state still leads to accurate XANES cross-sections: very similar spectra are obtained if the $2p_{\frac{3}{2}}$ radial wave-function from fully relativistic atomic calculations is employed as initial state instead of the $2p$ state from scalar relativistic calculations (see Figure S3 in the ESI\dag). 

\subsection{Computational Details\label{sub:computational-details}}

All electronic structure calculations have been performed with the QE distribution\cite{Giannozzi2009, Giannozzi2017}, using the PBE exchange-correlation functional\cite{Perdew1996, Perdew1997}. Pseudo-potentials from the GBRV library\cite{Garrity2014} and pslibrary\cite{DalCorso2014} have been employed, following the guidelines from the the standard solid-state pseudo-potential library (SSSP efficiency 1.0)\cite{Prandini2018}. In order to perform XANES cross-section simulations, the pseudo-potentials for oxygen and iridium have been generated using the atomic code in the QE distribution, including information to perform the reconstruction of the all-electron wave function. The electron wave-function and charge density have been expanded in plane waves with kinetic energy up to 50 Ry ($\sim 680$ eV) and 400 Ry ($\sim 5442$ eV), respectively. The first Brillouin zone of bulk IrO$_2$ has been sampled with a $\Gamma$-centered 8$\times$8$\times$12 k-point grid. Two-dimensional meshes with an equivalent spacing have been employed for the slab calculations. A cold smearing\cite{Marzari1999} with a width parameter of 0.01 Ry ($\sim0.136$ eV) has been applied. 

The electrochemical environment has been accounted for at a continuum level using the ENVIRON module for QE\cite{ENVIRON}. In particular, we have used the density-based self-consistent continuum solvation (SCCS) model in its original parametrization\cite{Andreussi2012}. A solvent-aware cavity\cite{Andreussi2019} has been employed to guarantee that small pockets in the quantum mechanical region (for instance, inside hydrogen-bonded OH networks) would remain dielectric free. The diffuse layer structure in the continuum solution has been modeled using the Poisson-Boltzmann (PB) model\cite{Nattino2019}. The analytical solution of the PB equation along the surface normal has been used to correct the electrostatic potential beyond the Stern layer, which we have set at 5 \AA\ from the outermost oxygen or iridium atom. Calculations have been performed for a 1:1 electrolyte with a bulk concentration of 0.1 M, which mimics the 0.1 M  H$_2$SO$_4$ solution employed in the \emph{operando} XANES  experiments.  

Slabs have been constructed using the suitable tools in the pymatgen library\cite{Ong2013}. Inversion symmetry has been enforced, so that all prepared surface systems exhibits two identical terminations. The surface structures have been constructed from the equilibrium bulk primitive cell, with lattice constants $a=b=4.520$ \AA\ and $c=3.196$ \AA. Geometry optimizations have been carried out in order to determine the equilibrium structure for each surface termination and applied charge. A maximum force of $2\times10^{-4}$ Ry/Bohr ($\sim 5$ meV/\AA) has been employed as convergence threshold. Frequencies have been computed from a standard finite-difference approach as implemented in the corresponding tool in the atomic simulation environment (ASE)\cite{HjorthLarsen2017}.

\section{Results and Discussion\label{sec:results}}

\subsection{XANES of Bulk IrO$_2$\label{sub:bulk}}

\begin{figure}
\centering
\includegraphics[width=\columnwidth]{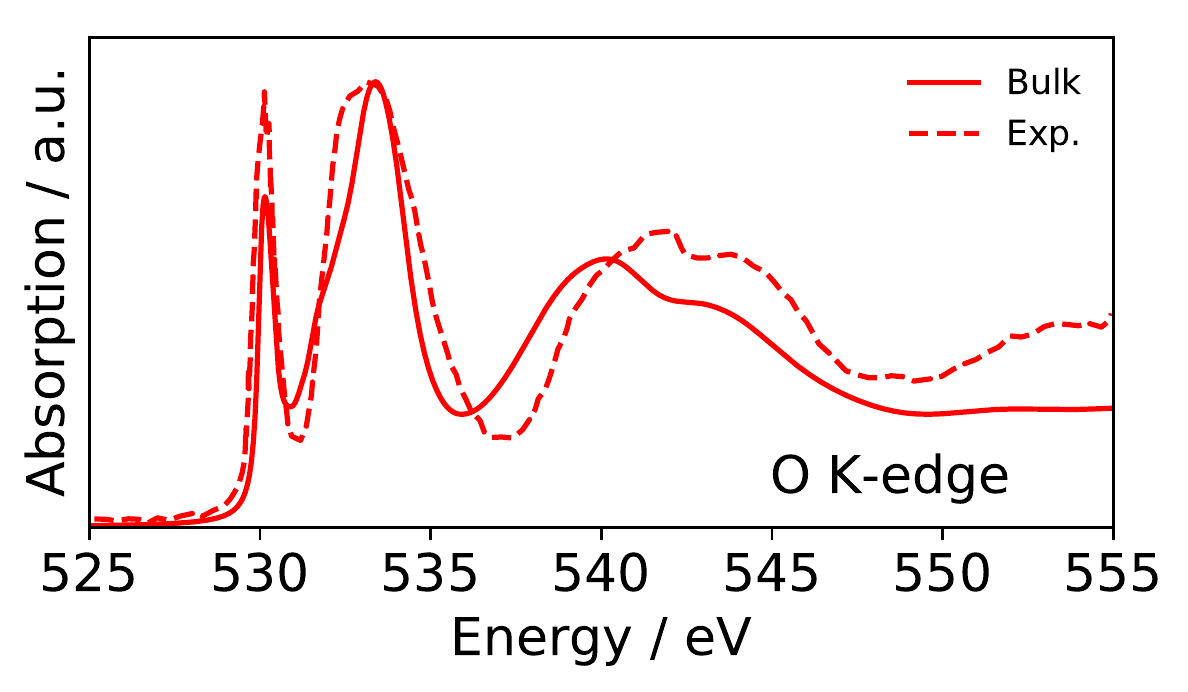}
\caption{Oxygen K-edge XANES absorption spectrum. Comparison of experimental data (dashed line)\cite{Pfeifer2016} on powder rutile IrO$_2$ and bulk simulations (solid line).}
\label{fig:OXASbulk}
\end{figure}

We start by considering the XANES cross sections computed for bulk IrO$_2$. Numerous experimental datasets have been computed for crystalline powder samples, looking at both the oxygen K-edge and at the iridium L$_3$/L$_2$ edges. Figure  \ref{fig:OXASbulk} illustrates the comparison between the experimentally measured spectrum at the O K-edge for a powder sample and the simulated spectrum. As already shown in Refs.\cite{Pfeifer2016, Pfeifer2016b}, the spectrum computed from the single-particle-based method also employed here agrees very closely with experimental data. As already discussed in Section \ref{sub:xanes}, the electron BE for bulk IrO$_2$ is set to 530 eV in order to best align the theoretical and experimental cross sections on the energy axis. All spectral features are well reproduced by the simulations: a narrow while-line peak, a broader peak within 5 eV from the absorption edge, and two broad peaks within 15 eV from the edge. Note that more elaborate approaches based on many-body perturbation theory\cite{Soininen2001, Vinson2011} have been shown to provide a similar level of agreement with experimental data\cite{Pfeifer2016a, Frevel2019}. 

\begin{figure}
\centering
\includegraphics[width=\columnwidth]{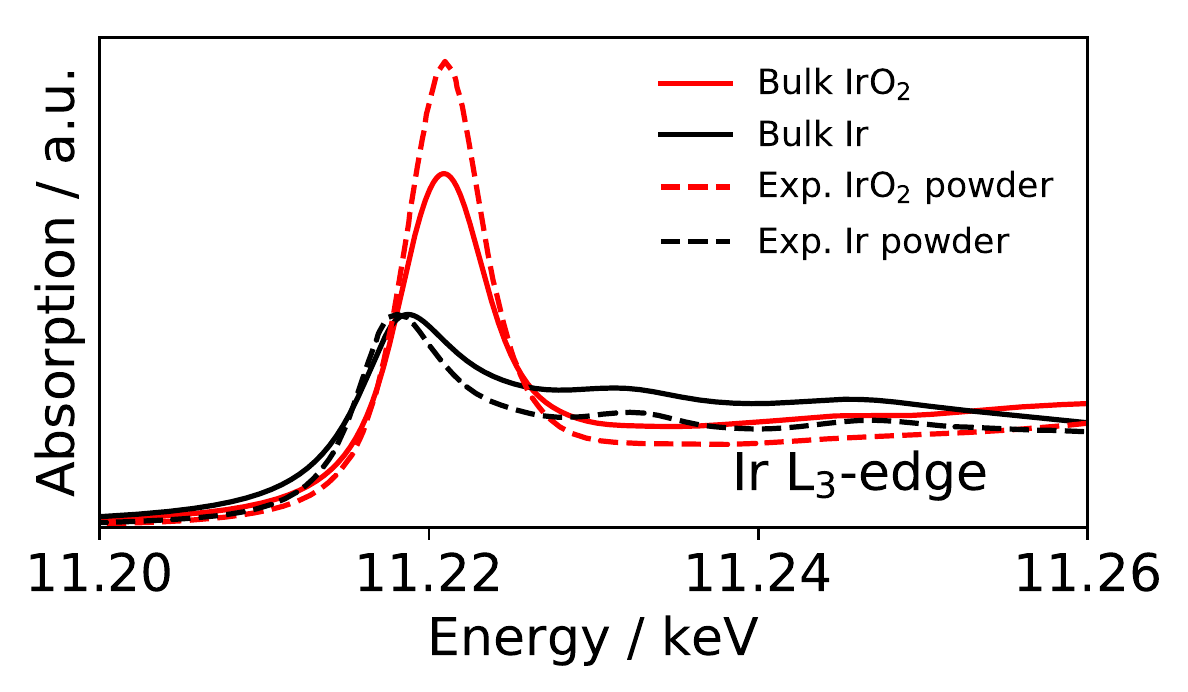}
\caption{Iridium L$_3$-edge XANES absorption spectrum. Comparison of experimental data on powder Ir (black dashed line) and IrO$_2$ (red dashed lines) crystals\cite{Clancy2012} and bulk simulations (solid lines, coloring as for the corresponding experimental system).}
\label{fig:IrXASbulk}
\end{figure}

Figure \ref{fig:IrXASbulk} illustrates the cross sections computed for IrO$_2$ at the Ir L$_3$-edge. Simulations are compared to experimental data on crystalline powder samples, as determined by Clancy et al.\cite{Clancy2012}. Similarly to what carried out for the oxygen K-edge, we set the reference BE of the Ir $2p$ electron in IrO$_2$ to best align the theoretical cross-section (solid red line in Figure \ref{fig:IrXASbulk}) to the experimental one (dashed red line in Figure \ref{fig:IrXASbulk}). In order to test the sensitivity of the XANES simulations to the oxidation state of the Ir atoms, we have additionally computed the absorption spectrum for (bulk) elemental Ir. The simulated spectrum is also presented in Figure \ref{fig:IrXASbulk} together with the experimentally-determined cross section\cite{Clancy2012}. Note that the energy shift between the calculated Ir and IrO$_2$ cross sections is fully based on results of first-principles calculations: it corresponds to the $2p$-electron BE difference in the two materials. It is only the absolute position of the IrO$_2$ spectrum on the energy axis that involves a free parameter. Simulations correctly reproduce both the direction and the magnitude of the shift of the while-line peak that is observed for a decrease in the Ir oxidation state (from $+4$ in IrO$_2$ to $0$ in elemental Ir). The computed spectra also qualitatively reproduce the relative intensity of the two white-line peaks, with the absorption edge of IrO$_2$ having the largest intensity. However, simulations underestimate the relative intensity of the two peaks, which might be due to the neglect of the SOC in the final-state calculation. 

Overall, Figure \ref{fig:OXASbulk} and Figure \ref{fig:IrXASbulk} show that the single-particle-based approach from Refs.\cite{Taillefumier2002, Gougoussis2009a, Bunau2013} is able to predict accurate XANES cross sections for IrO$_2$ both at the oxygen K-edge (as already shown in Refs.\cite{Pfeifer2016, Pfeifer2016b}) but also at the iridium L$_3$-edge. 

\subsection{IrO$_2$ Interfaces: Electrochemical Stability\label{sub:surf-stability}}

In the search of the most stable IrO$_2$ interfaces, we have considered surfaces from five low-index crystallographic planes: (111), (110), (101) (or (011)), (100) (or (010)) and (001). For each plane, all inequivalent terminations have been considered. For all surfaces presenting unsaturated oxygen atoms, we have additionally included in the analysis hydroxy-terminated interfaces. In particular, we have considered surfaces where bridge-site oxygen atoms have been hydrogenated, surfaces where oxygen atoms at the coordinatively-unsaturated (CU) sites have been hydrogenated, and surfaces where both bridge- and CU-oxygen atoms have been hydrogenated. For the surfaces with unsaturated iridium atoms, we have considered both hydroxy- and oxygen-terminated interfaces in addition to the clean surfaces. Overall, 37 IrO$_2$ surface configurations have been considered.  

\begin{figure}
\centering
\includegraphics[width=0.55\columnwidth]{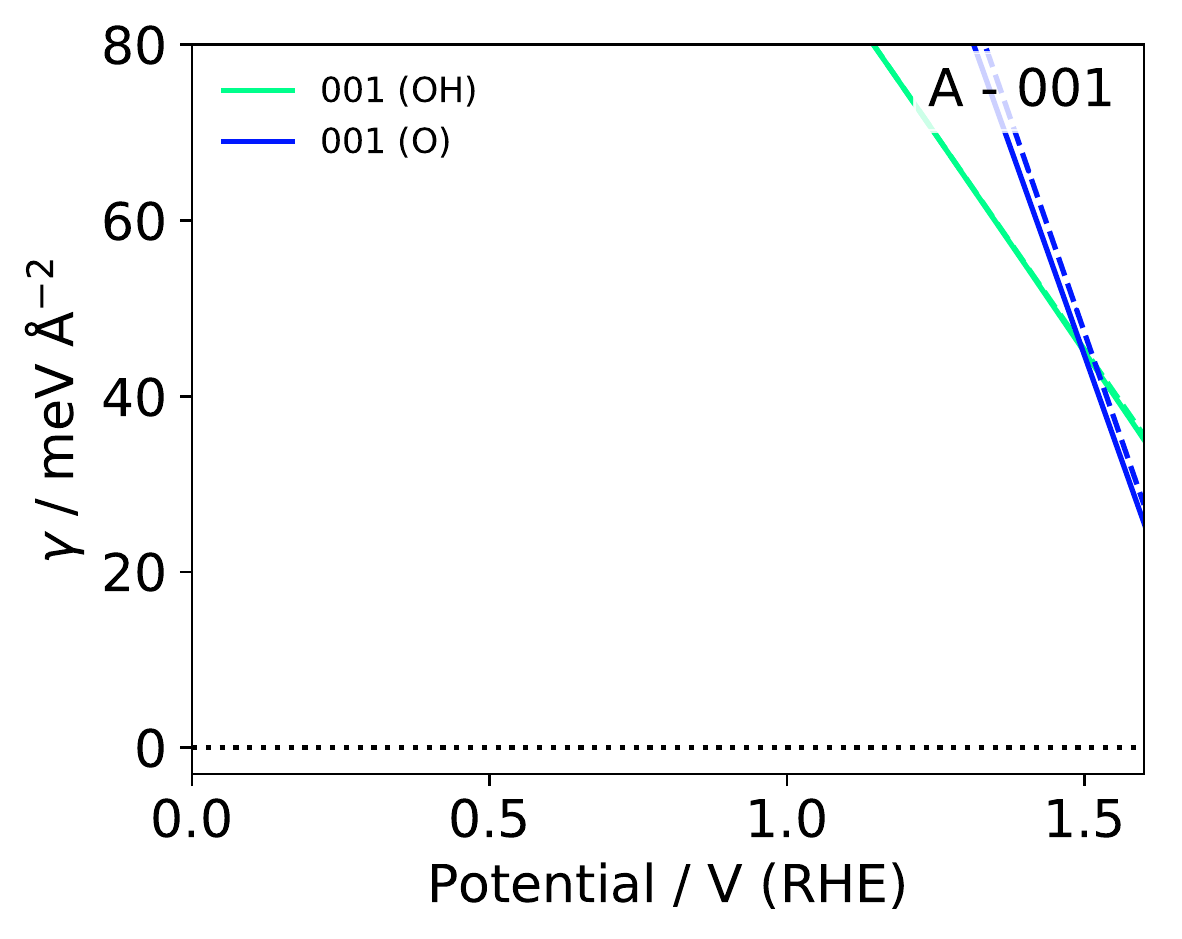}\\
\includegraphics[width=0.55\columnwidth]{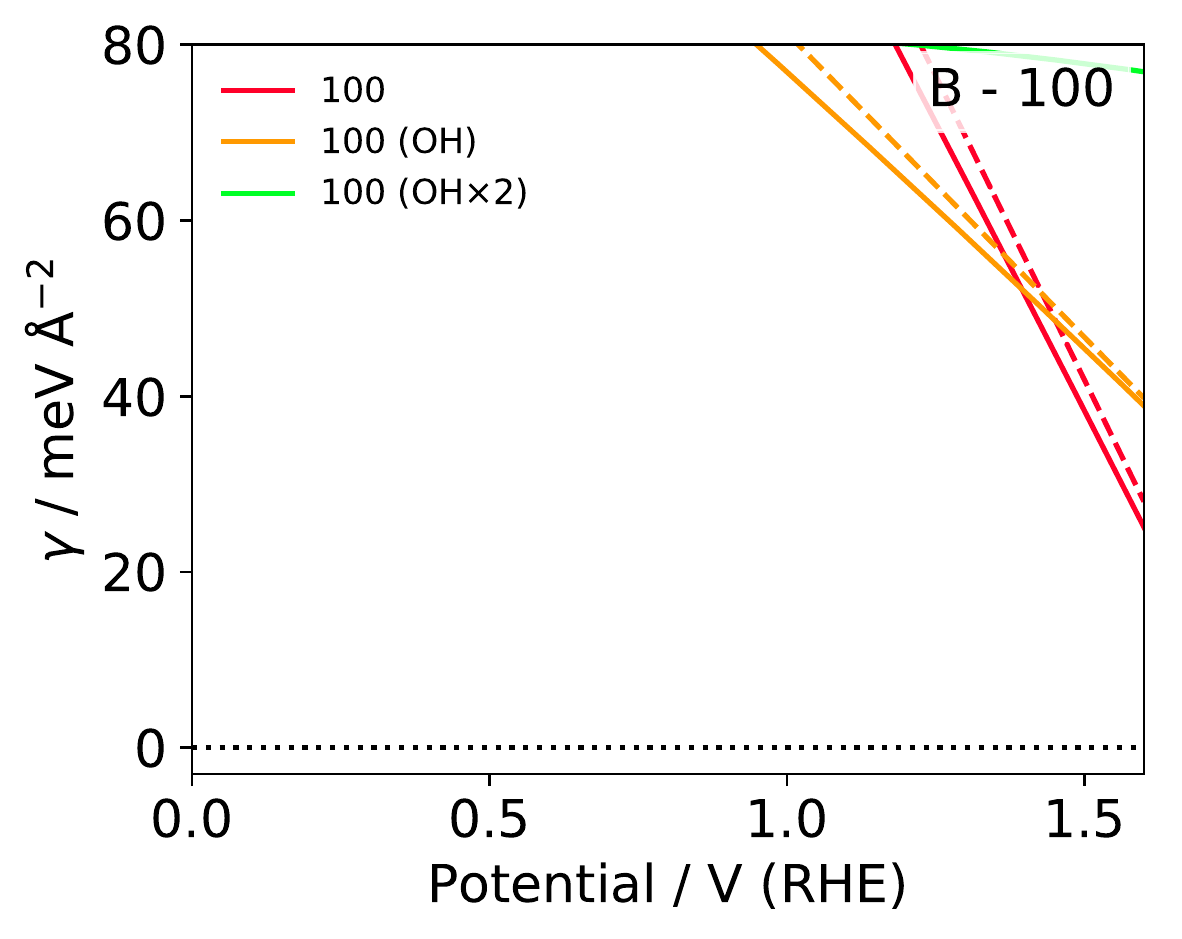}\\
\includegraphics[width=0.55\columnwidth]{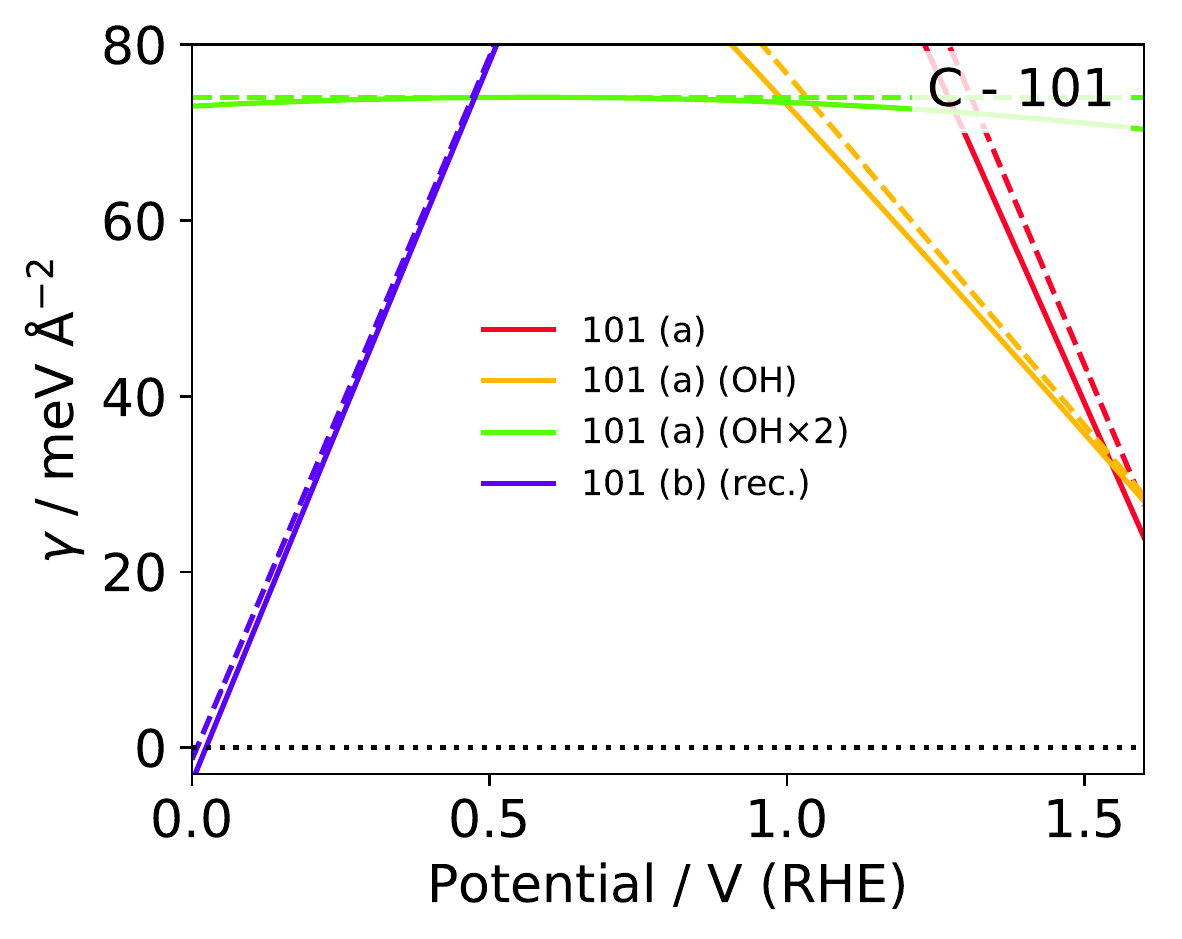}\\
\includegraphics[width=0.55\columnwidth]{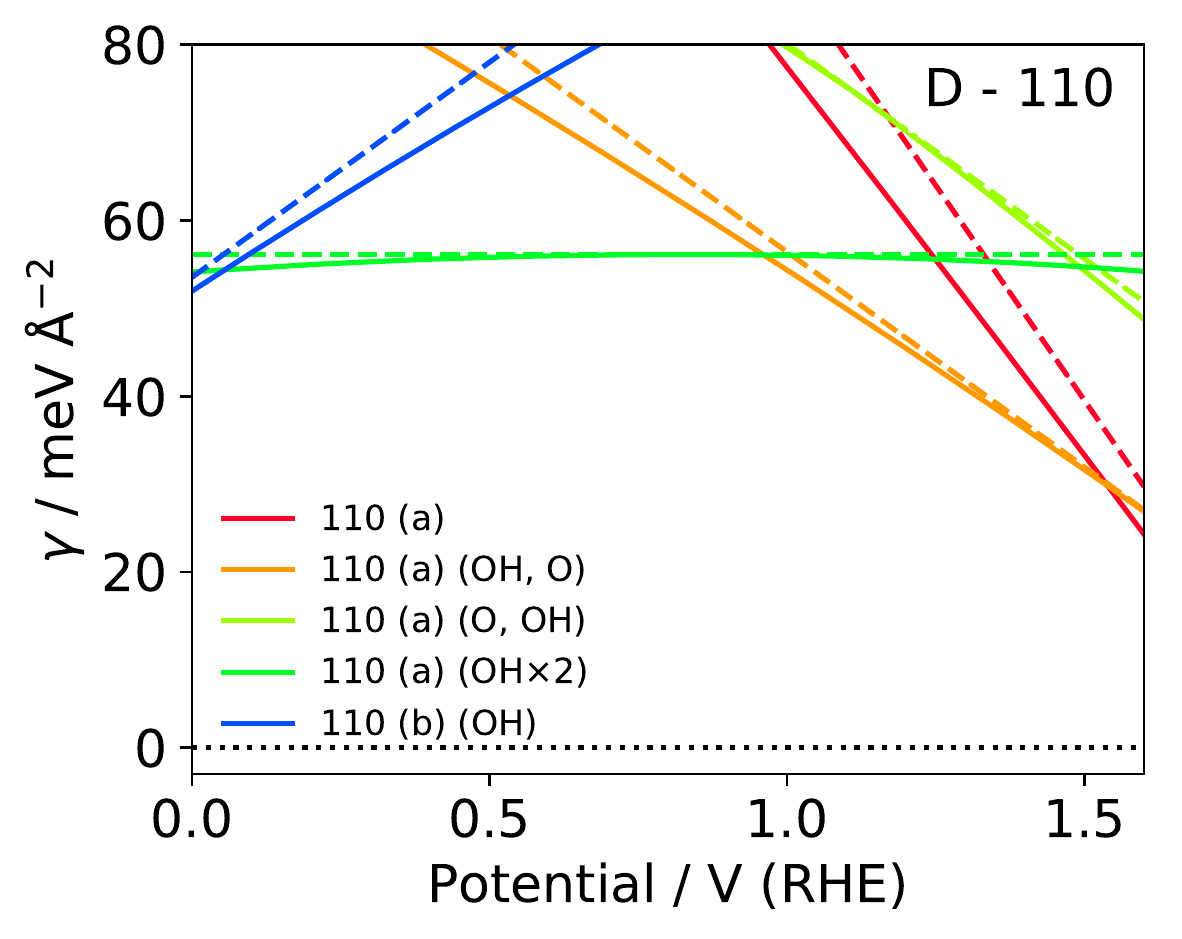}\\
\includegraphics[width=0.55\columnwidth]{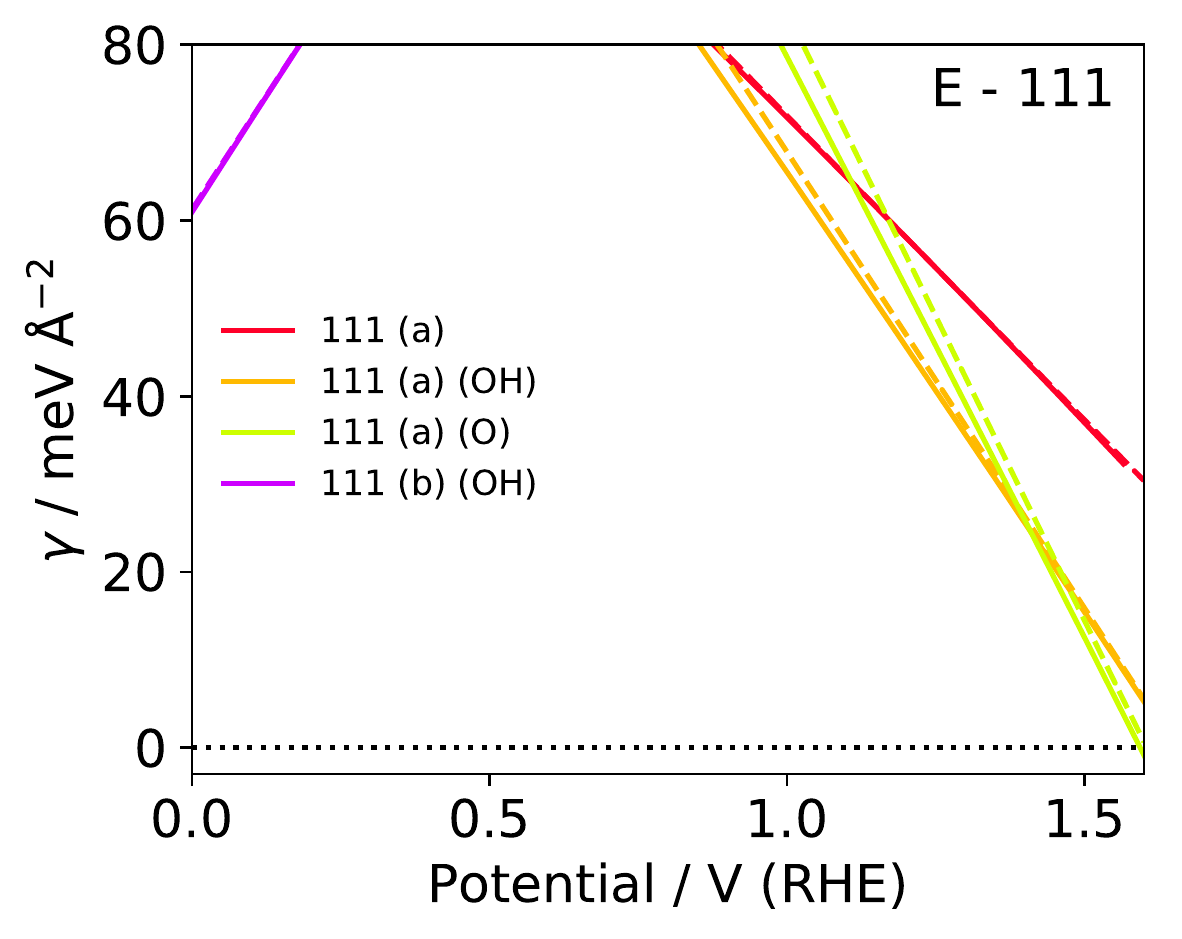}\\
\caption{Surface energy as a function of the applied potential for the considered terminations: 001 (A), 100 (B), 101 (C), 110 (D) and 111 (E). Only the interfaces with surface energy lower than 80 meV/\AA\ for a potential value in the range $0<U<1.6$ have been illustrated. Dashed lines illustrate surface energies computed using the CHE approach, while solid lines illustrate results obtained using a fully grand-canonical method. A pH of 1 has been considered for the grand-canonical interface energies in order to approximately accounting for the 0.1 M  H$_2$SO$_4$ electrolyte solution employed in the\emph{operando} XANES  experiments.}
\label{fig:stability}
\end{figure}

Results of the stability analysis are illustrated in Figure \ref{fig:stability}, where the surface free energies of the most stable interfaces are plotted as a function of the applied potential for each of the terminations considered. For each interface configuration, Figure \ref{fig:stability} A-E include results obtained using the CHE approach (straight dashed segments) as well as with a full grand-canonical method (curved solid lines). As discussed in Ref.\cite{Hormann2019}, the surface free-energies obtained with the CHE represent upper bounds for the full grand-canonical surface energies. At the potential of zero charge (PZC) of each interface, the two approaches predict identical surface energies. For charged interfaces, in the full grand-canonical approach the deviation from the CHE is close to quadratic with the applied potential, as it arises from the explicit inclusion of a double-layer-related capacitor-like electrostatic energy contribution. Despite quantitative differences for potentials that significantly deviate from the PZC, we find that the two approaches predict very similar stability curves, which lead to the same qualitative trends concerning the relative stability of the various interfaces. In the following, we will report potentials at which transitions between interfaces occur using the more accurate grand-canonical approach.

The (101) and (110) terminations have been found to have lowest surface energies for low-potential conditions (close to 0 V). In particular, the (110) surface is the minimum-energy low-index termination at zero temperature\cite{Ping2015}, and it has been suggested to be the most stable surface at finite temperature under low potential condition\cite{Matz2017, Opalka2019}. Consistently with previous studies\cite{Ping2017, Pfeifer2017, Opalka2019}, we find the fully-hydrogenated surface to be the most stable configuration for the (110) surface close to zero potential (see Figure  \ref{fig:stability} D). Deprotonation of the hydroxyl groups takes place at higher potentials, for the OH groups at the bridge sites (at $\sim$1 V) and at the CU sites (at $\sim$1.5 V). 

Interestingly, we find a reconstruction of the iridium-rich (101) termination to be the minimum-energy interface for low potential conditions (see Figure  \ref{fig:stability} C). Oxygen atoms that are sub-surface in the bulk-terminated configuration are found to emerge to the  top layer, leading to under-coordinated iridium atoms in the first two layers and an inter-layer oxygen vacancy. An illustration of the bulk-terminated surface and the reconstructed interface is provided in the ESI\dag (see Figure S4). Note that substantial surface reconstruction accompanied by modification of the coordination shell of first-layer iridium atoms has been already observed for open surfaces and oxygen-poor conditions\cite{Pfeifer2016a}. For the reconstruction considered, we find negative surface energies for potentials that are lower than $\sim 0$ V, which would suggest a significant driving force for the formation of similarly-reconstructed IrO$_2$ surfaces close to open-circuit conditions. 

The reconstruction is predicted to be lifted at about 0.5 V, at which potential the surface energy of the OH-covered (101) interface resembles the one of the OH-covered (110) surface, with hydroxyl groups at both bridge and CU sites. Also, similarly to the (110) case, the OH groups at bridge sites first deprotonate at $\sim$1 V, followed by the ones at the CU sites at $\sim$1.5V.  

In agreement with the findings from Opalka \emph{et al.}\cite{Opalka2019}, the (111) surface is found to be the most stable termination at large applied potentials, and the first interface to present a negative surface energy (see Figure  \ref{fig:stability} E). In particular, for $U > 1.2$ V, a (111) termination with a hydroxyl group adsorbed on top of the CU iridium atoms becomes the overall minimum-energy interface. The OH group is then deprotonated at approximately $1.4$ V, with the resulting interface energy becoming negative at $\sim$1.6 V. At potentials lower than 0.9 V, the hydroxyl adsorbed at the CU site desorbs, and at even lower potentials the bridge oxygen atoms are protonated. Finally, the (001) and (100) terminations (Figure  \ref{fig:stability} A and B) are found to be overall the less stable interfaces, with large surface energies under both low- and high-potential conditions.  

From the experimental point of view, \emph{operando} OER investigations of iridium-oxide-based catalysts mostly focused on electrochemically-oxidized Ir nanoparticles\cite{Pfeifer2016, Pfeifer2016b, Pfeifer2016a, Saveleva2018, Frevel2019}, where the outer oxide layer is characterized by an amorphous crystallographic form (IrO$_x$). Such phase of iridium oxide has been shown to exhibit higher activity towards oxygen evolution as compared to crystalline rutile IrO$_2$ nanoparticles\cite{Reier2014, Pfeifer2016b, Saveleva2018}. Low-index rutile IrO$_2$ surfaces, which best suit periodic DFT simulations, are instead considered by theory, limiting the possibility of a direct comparison with experiments. Nevertheless, crystalline IrO$_2$ nanoparticles obtained from thermal synthetic routes\cite{Saveleva2018, Abbott2016, Povia2019} show a predominance of the (101) and (110) surfaces\cite{Abbott2016, Povia2019}. While the ratio between the two terminations depends on the exact conditions, the abundance of these surfaces on thermally-prepared nanoparticles is consistent with the prediction of these terminations having the lowest-energy for open-circuit conditions. Note that despite differences with the experimentally-considered amorphous IrO$_x$, the investigation of the most stable rutile IrO$_2$ interfaces under electrochemical conditions remains meaningful as a test for the electrochemical stability of local coordination environments and adsorbed species that might play a role as active sites on the catalyst surface. 

\subsection{\emph{Operando} XANES of IrO$_2$ Interfaces\label{sub:XANES-results}}

After having considered the electrochemical stability of the various IrO$_2$ terminations we turn to the simulation of  X-ray absorption spectra for selected interfaces under the effect of applied potential. For the structures employed for the calculation of the XANES cross-section, we use the same relaxed interface configurations that we have employed for the estimate of the stability curves. Explicit charges are included to directly mimic the effect of the applied voltage, and the continuum solvent and electrolyte media account for the presence of the electrochemical environment. In this way, we can study potential-dependent structural changes and solvent- and electrolyte-induced polarization and charge-stabilization effects. Note that atomic reorganizations and potential-induced surface charge accumulation only affect the first few atomic layers, and we consistently find essentially bulk-like XANES cross sections for absorbing atoms from the second/third layer inwards.

\subsubsection{Oxygen K-edge\label{subsub:OKedge}}

\begin{figure}
\centering
\includegraphics[width=\columnwidth]{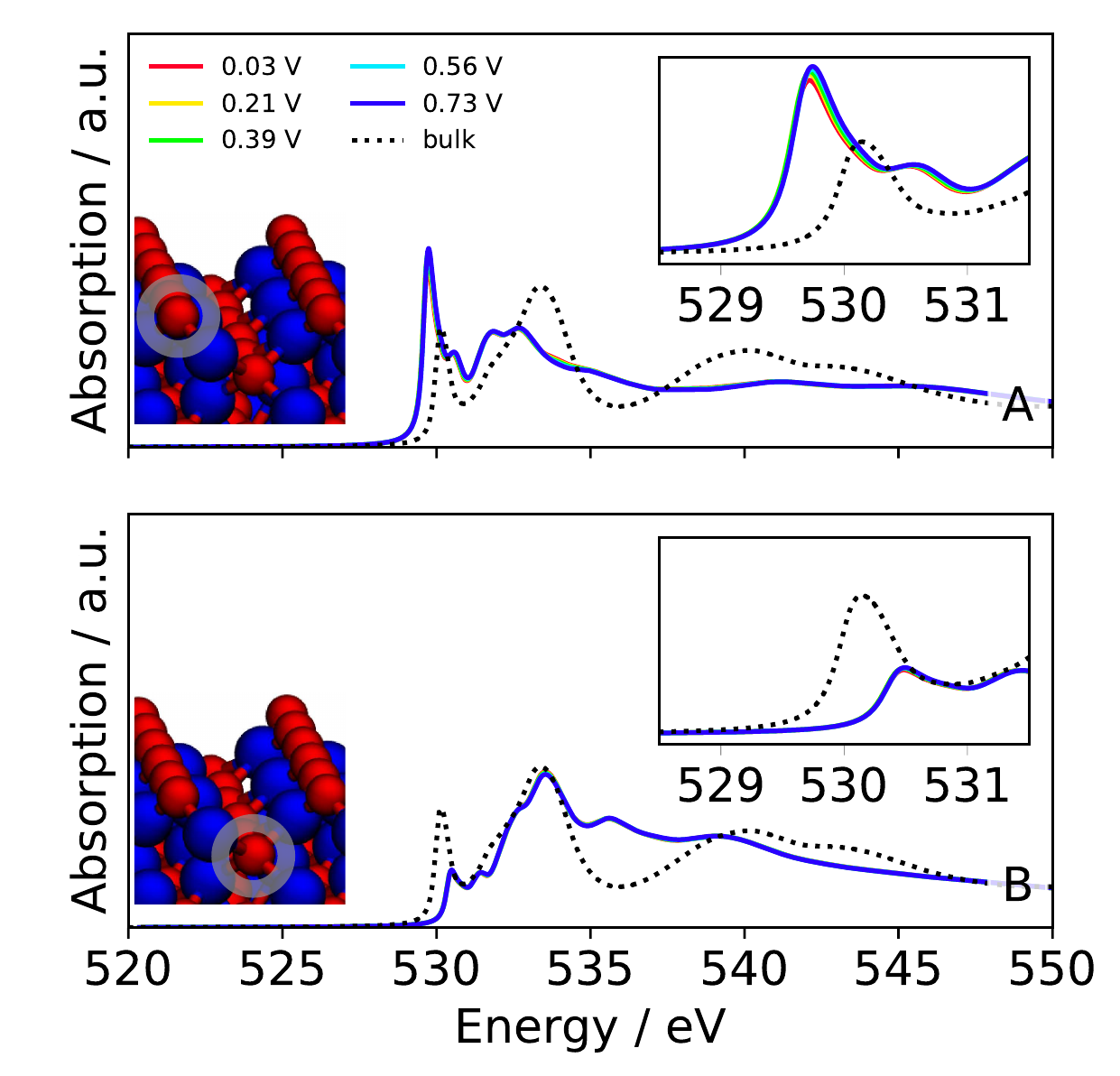}\\
\caption{Oxygen K-edge XANES cross sections computed for the reconstructed iridium-rich (101) surface. A sketch of the termination is presented in the inset on the left, with blue and red balls indicating iridium and oxygen atoms, respectively. The absorbing atoms are highlighted with a grey circle. The insets on the right include a magnification of the white-line peak region. The various colors identify different potential conditions (curves are to a large extend superimposed). }
\label{fig:101-rec}
\end{figure}
\begin{figure}
\centering
\includegraphics[width=\columnwidth]{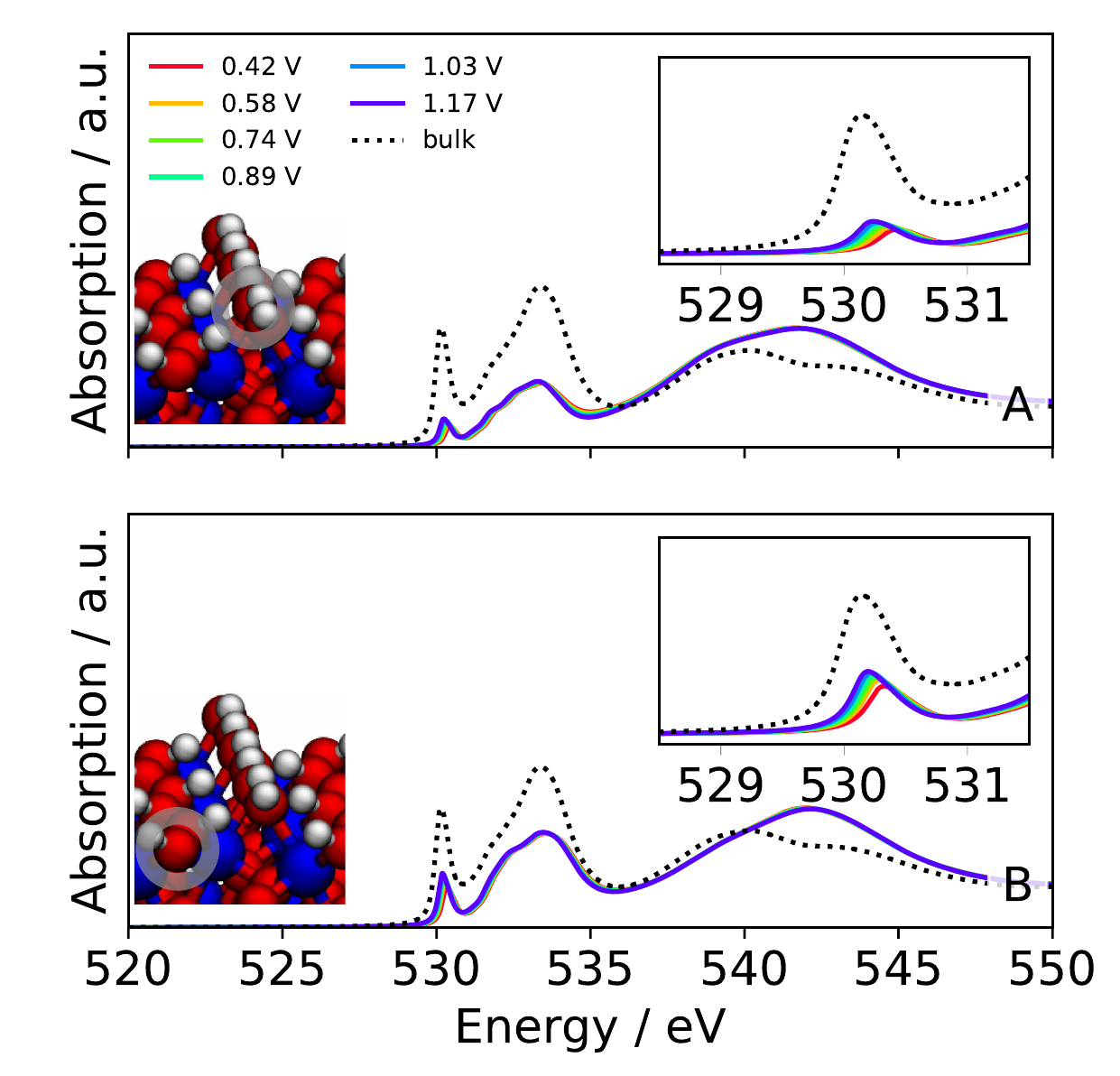}\\
\caption{Same as Figure \ref{fig:101-rec}, but the absorbing atoms are selected from the OH-covered (101) surface.}
\label{fig:101-Hall}
\end{figure}

We start by looking at the spectra computed at the oxygen K-edge, and we first consider the (101) termination, focusing on the interface that we have found to be most stable under low potential conditions (see Figure \ref{fig:stability}C). Figure \ref{fig:101-rec} illustrates potential-dependent XANES cross sections as computed for the symmetry inequivalent atoms in the first two layers of the reconstructed Ir-rich termination considered. The white-line peaks corresponding to the two absorbing atoms are found to differ from the ones computed in bulk IrO$_2$ both in position and intensity.  The absorption edge of the bridge oxygen atom in the top-most layer is only slightly shifted to lower energies by less than 0.5 eV. A similar shift in the opposite direction is observed for the second-layer atom. Note that the potential-dependence observed for the two absorbing atoms is essentially negligible, with a somewhat larger effect for the first-layer atom, where the surface charge is more likely to localize. 

At about $\sim$0.5 V the surface reconstruction is predicted to be lifted, and the OH-covered interface becomes the most stable (101) termination (see Figure  \ref{fig:stability}C). Figure \ref{fig:101-Hall} illustrates the XANES cross sections computed for the O atoms in the two symmetry-inequivalent hydroxyl groups in the first layer. Interestingly, the white-line peaks for both oxygens lie very close to the oxygen peak in bulk IrO$_2$. Their intensity, however, is much smaller. Similarly, the broader peak at $\sim$533 eV is less intense in the spectra of the hydroxyl groups, and a new feature is present at $\sim$543 eV. The increasing potential strengthens the OH binding, weakening the hydrogen-bond network and shifting the absorption edges towards lower energies. 

The change in the absorption edge position from the reconstructed (101) surface to the OH-covered interface, which should take place at $\sim0.5$ V, is predicted to be relatively small (i.e. of the order of $0.5$ eV). This is maybe why \emph{operando} XANES investigations of rutile IrO$_2$ nanoparticles have not observed detectable changes in the white-line position at the oxygen K-edge from 0 V to 0.7 V, even though a 0.5 V peak has been observed in the cyclic voltammetry\cite{Saveleva2018}. No major changes in the absorption edge were also observed for electrochemically-oxidized Ir nanoparticles, which, however, are constituted by an amorphous phase of iridium oxide (IrO$_x$)\cite{Pfeifer2016a, Saveleva2018, Frevel2019}. 

\begin{figure}
\centering
\includegraphics[width=\columnwidth]{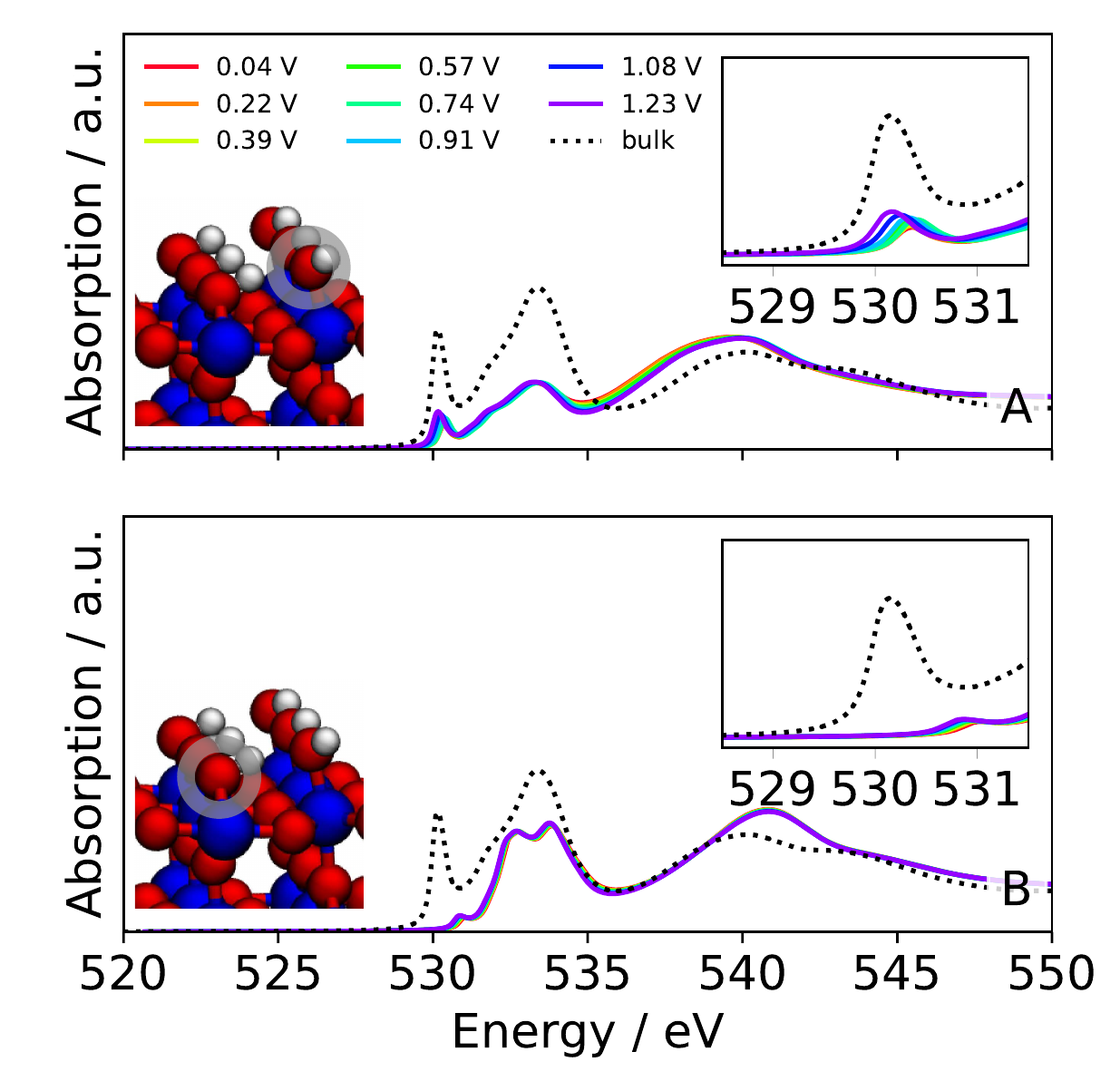}\\
\caption{Same as Figure \ref{fig:101-rec}, but the absorbing atoms are selected from the OH-covered (110) surface.}
\label{fig:110-Hall}
\end{figure}

Similar cross-sections to the ones predicted for the hydroxylated (101) termination (Figure \ref{fig:101-Hall}) are obtained for the analogous OH-covered (110) interface, see Figure \ref{fig:110-Hall}. The (110) termination presents oxygen atoms at bridge and CU sites as well, both occupied by hydroxyl groups for potentials up to 1 V (see Figure \ref{fig:stability}D). The XANES spectrum for the OH group at the CU site (Figure \ref{fig:110-Hall}A) very much resembles the spectrum computed for the same site on the (101) surface (cf. Figure \ref{fig:101-Hall}A). For what concerns the OH group at the bridge site, the (110) white-line peak falls 1 eV higher in energy than the corresponding peak for the (101) surface. Note that the absorption edge is sensitive to the oxidation state of the absorbing atoms, with the oxygen edge shifting to lower energies for its increasing electron deficiency\cite{Saveleva2018}. Thus, despite the similarity of the two terminations, the bridge-site hydroxyl groups on the two surfaces present a different local charge state, with a higher electron deficiency on the bridge oxygen of the (101) termination. This is consistent with the higher OH-density on  this surface, with both CU- and bridge-site hydroxyl groups being coordinated to the same iridium atom.

\begin{figure}
\centering
\includegraphics[width=\columnwidth]{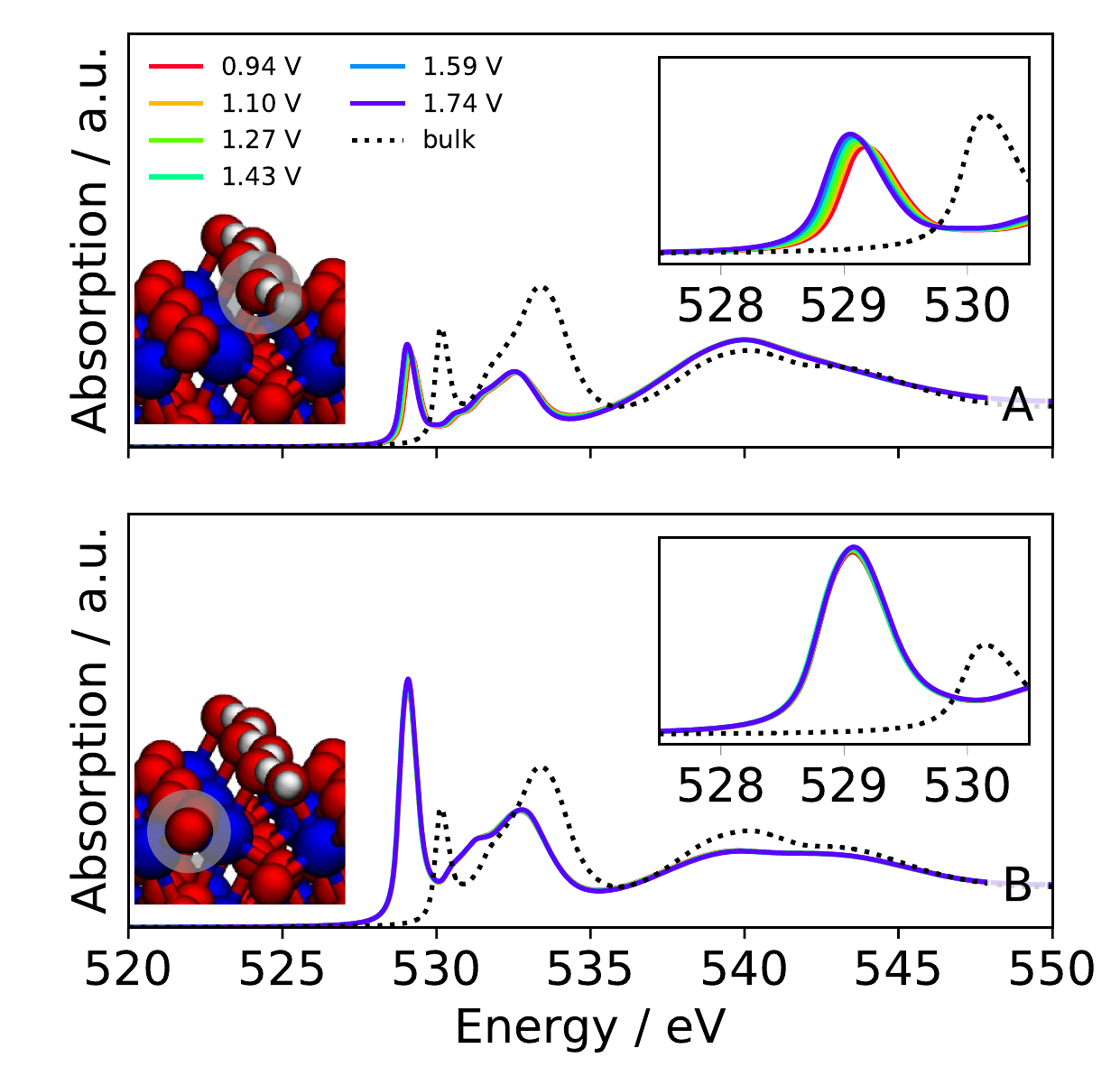}\\
\caption{Same as Figure \ref{fig:101-rec}, but the absorbing atoms are selected from a (101) termination with OH groups at the CU sites.}
\label{fig:101-H1}
\end{figure}
\begin{figure}
\centering
\includegraphics[width=\columnwidth]{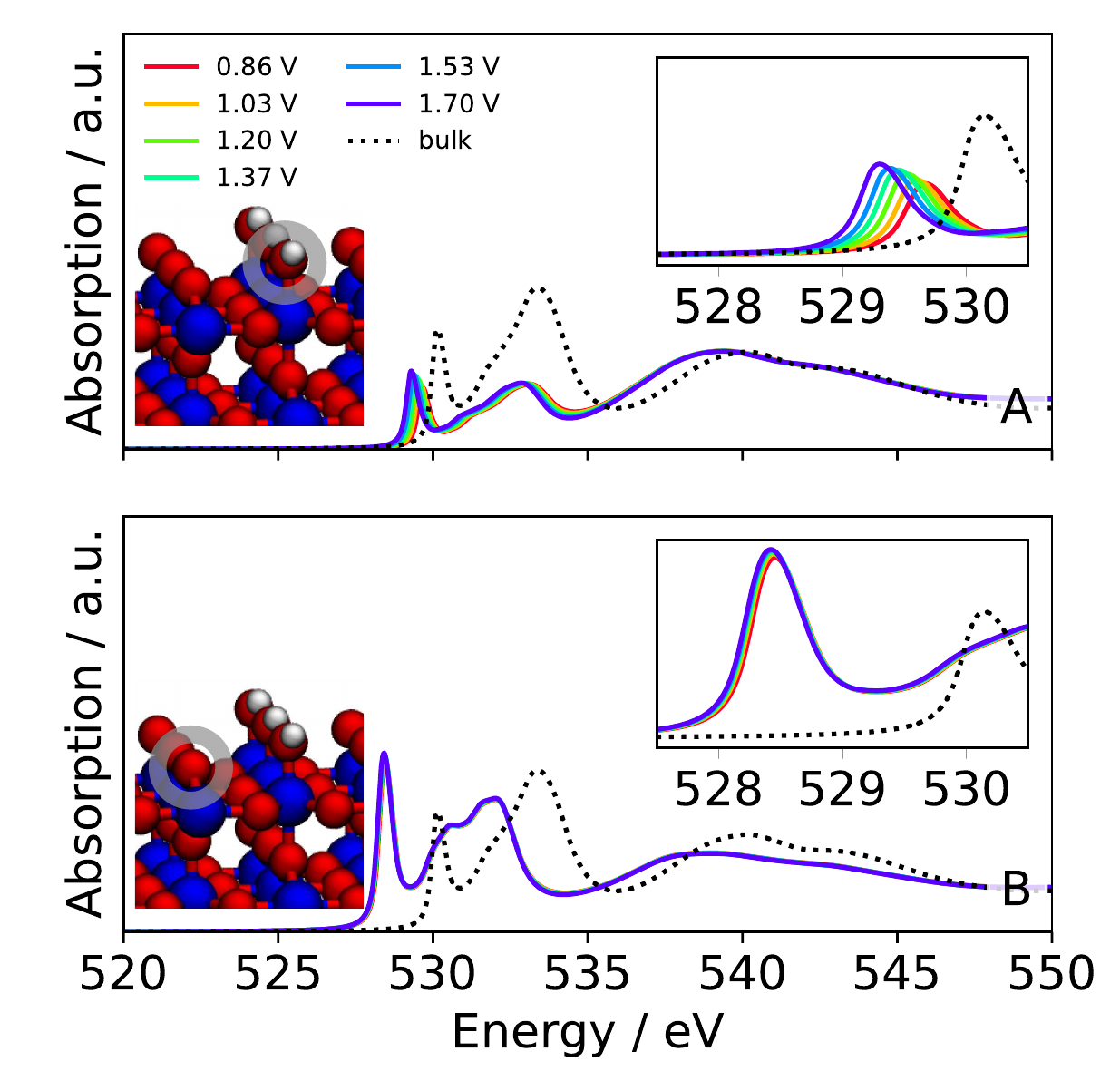}\\
\caption{Same as Figure \ref{fig:101-rec}, but the absorbing atoms are selected from a (110) termination with OH groups at the CU sites.}
\label{fig:110-H1}
\end{figure}

Figure \ref{fig:101-H1} and Figure \ref{fig:110-H1} illustrate the cross sections computed for the first-layer oxygen atoms after the deprotonation of the bridge-site OH groups on the (101) and the (110) terminations, respectively. These interfaces are predicted to become the most stable interfaces at $\sim1$ V. For the (101) surface (Figure \ref{fig:101-H1}), the white-line for both absorbing oxygens peaks at about 529 eV. A similar absorption edge has been shown\cite{Pfeifer2016, Pfeifer2016b, Pfeifer2016a, Pfeifer2017, Frevel2019} to correspond to an oxygen in a formally $-1$ oxidation state (O$^{\mathrm{I}-}$), with holes in the 2p states. Bulk iridium vacancies\cite{Pfeifer2016, Pfeifer2016b} and under-coordinating surface environments\cite{Pfeifer2016a, Saveleva2018, Frevel2019} have been shown to give rise to similar features. The predicted deprotonation of the bridge-site oxygens (sometimes referred to as $\mu_2-O$'s) and the simulated absorption edge at 529 eV for the O and OH surface species for potentials larger than 1 V are consistent with the observation of the appearance and the growth of an analogous feature in the experimental spectra\cite{Pfeifer2016a, Saveleva2018, Frevel2019}. 

The similar position of the white-line peaks for the two absorbing atoms in the partially-deprotonated (101) surface indicates that the two oxygens present a similar oxidation state despite the different chemical environment. On the (110) surface (Figure \ref{fig:110-H1}),  the absorption edge of the bridge-site O is at significantly lower energies than the one of the CU-site OH, suggesting a more significant hole localization on the bridge oxygen. For this surface, an evident potential dependence of the absorption spectrum is evident, with a $>0.5$ eV down-shift of the adsorption edge for a potential increase of $\sim$0.8 V (Figure  \ref{fig:110-H1}A). At the largest potentials considered (1.7 V), the position of the absorption edge is close to 529 eV. Virtually no potential dependence is instead observed for the bridge-site oxygen, with a white-line peak at $\sim$528.5 eV. 

\begin{figure}
\centering
\includegraphics[width=\columnwidth]{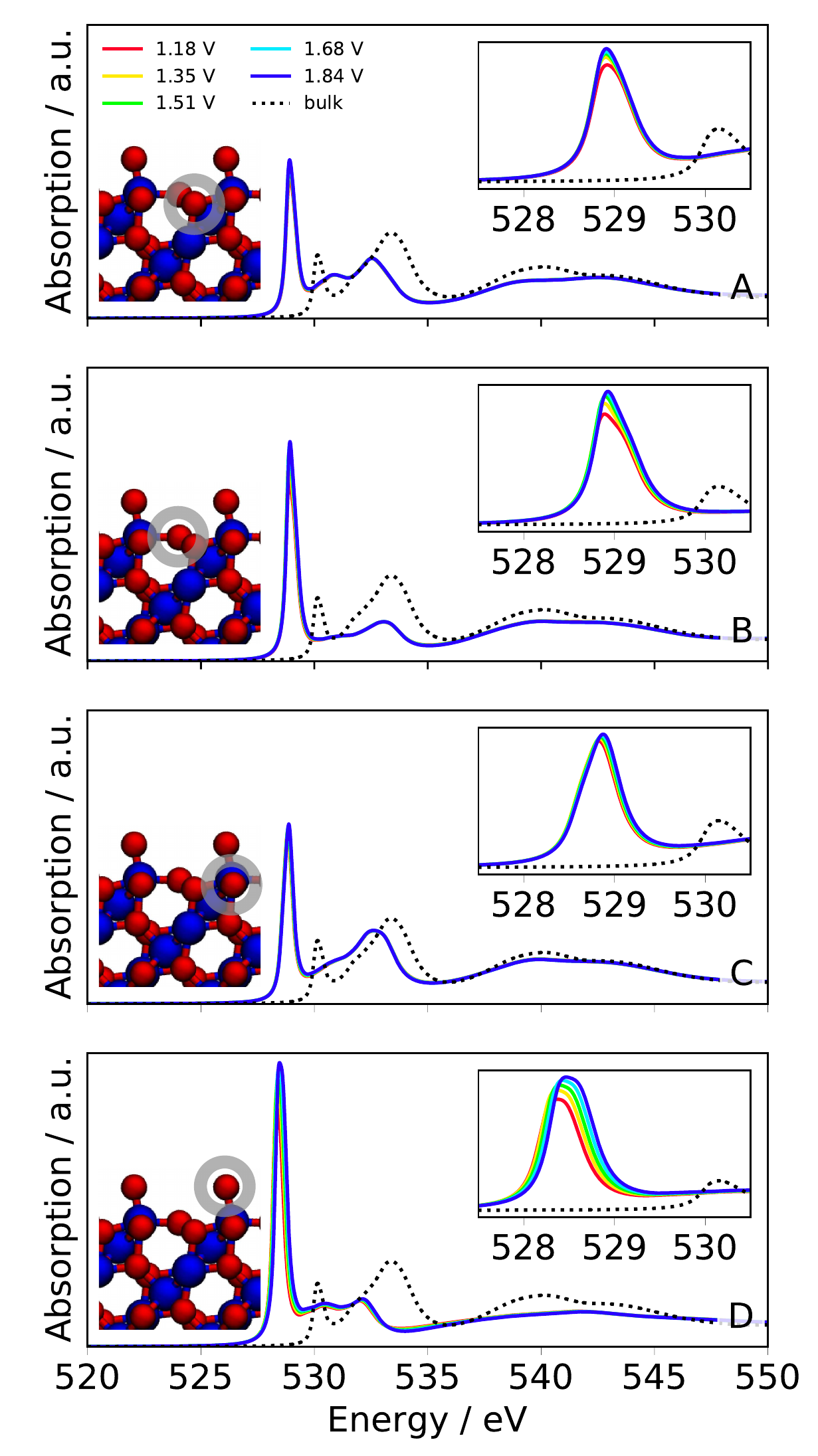}\\
\caption{Same as Figure \ref{fig:101-rec}, but the absorbing atoms are selected from a (111) termination with an oxygen atom adsorbed at the CU site.}
\label{fig:111-OX}
\end{figure}

For high potential conditions, CU-site oxygen atoms are predicted to form on the (101) and the (110) terminations from the deprotonation of the corresponding OH groups. This is predicted to take place at $\sim1.5$ V (see Figure \ref{fig:stability}C and D). For such conditions, however, it is the (111) termination with an oxygen atom at the CU site to present the lowest surface energy, which is why we consider this interface to illustrate the predicted XANES cross sections (Figure \ref{fig:111-OX}) while reporting the spectra for fully-deprotonated (110) and (101) interfaces in the ESI\dag (Figure S5 and S6, respectively). In addition to the CU-site oxygen, we consider the other symmetry inequivalent oxygen atoms that are present in the top layer of the surface, which are three bridge-site oxygens. For the three bridge oxygen atoms we observe only slight differences in terms of intensity for the features of the spectrum up to 535 eV (Figure \ref{fig:111-OX} A-C). The white lines for the three atoms also peak at very similar energies as the bridge-oxygen absorption edges that are observed for the (110) and (101) surfaces (close to $529$ eV, cf. Figure \ref{fig:101-H1} and \ref{fig:110-H1}). The absorption edge of the oxygen atom adsorbed at the CU site (also known as $\mu_1-O$) is instead located at a somewhat lower energy ($\sim528.5$ eV). The CU-oxygen white-line peak (Figure \ref{fig:111-OX}D) has the highest intensity and it exhibits the strongest potential dependence. Note that this atom has the lowest coordination number and its protrusion in the large-field region of the double-layer likely makes it the most sensitive to charge accumulation on the surface. Rather similar absorption edges have been found for the CU-site oxygens on the (101) and the (110) terminations, at about 528-527.5 eV, respectively (see Figure S5 and S6 in the ESI\dag). The absorption spectra for the bridge atoms on these surfaces are essentially identical to the ones predicted for the surfaces where the hydroxyl group at the CU-site is protonated (see Figures \ref{fig:101-H1} and \ref{fig:110-H1}). Overall, the prediction of the CU-site OH deprotonation at $\sim1.5$ V together with the suggested appearance of an absorption peak at about 528 eV are consistent with the findings of Frevel et al.\cite{Frevel2019}. They have reported the occurrence of an oxidation wave at 1.4 V on IrO$_x$ and the following appearance of a 528 eV peak in the measured \emph{operando} XANES cross-sections\cite{Frevel2019}. Our calculations thus agree with the simulations from Ref.\cite{Frevel2019} in the assignment of such oxidation wave to the formation of CU-site oxygens.

\subsubsection{Iridium L$_3$-edge\label{subsub:IrL3edge}}

\begin{figure}
\centering
\includegraphics[width=\columnwidth]{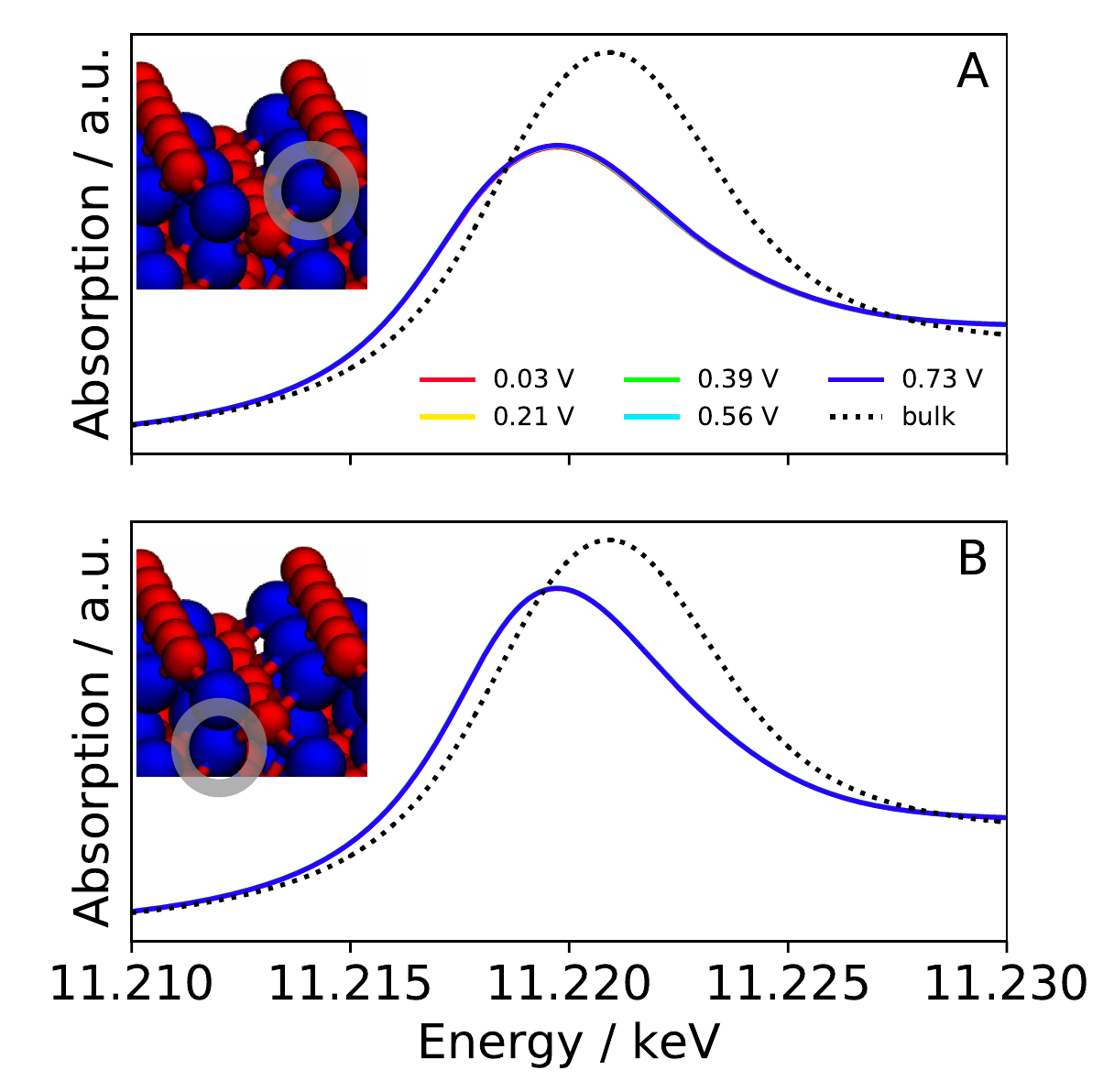}\\
\caption{Iridium L$_3$-edge XANES cross sections computed for the indicated absorbing atoms of the reconstructed iridium-rich (101) surface. The various colors identify different potential conditions (curves are to a large extend superimposed).}
\label{fig:Ir-101-rec}
\end{figure}

We illustrate the potential dependence of the XANES cross-section at the iridium L$_3$-edge using the (101) termination. As for the oxygen K-edge, we start by considering the reconstructed Ir-rich termination that is most stable for potentials close to 0 V (see Figure \ref{fig:stability}C). As absorbing atoms, we consider the two inequivalent Ir atoms in the first and second atomic layers, which are the ones that are mostly involved in the reconstruction. These atoms, whose XANES cross-sections are presented in Figure \ref{fig:Ir-101-rec}, have a significantly different chemical environment as compared to bulk atoms, with a reduced number of coordinating oxygen atoms (4 vs 6). Given the oxygen deficiency in the top-most layers, we expect a formally lower oxidation state for these iridium atoms. We consistently observe absorption edges that considerably differ from the ones computed for bulk atoms, with a significant shift of the order of 1 eV towards lower energies. In addition, the white-line peaks for the investigated absorbing Ir atoms are characterized by lower intensities, and this is especially true for the first-layer atom. These findings are consistent with the experimental XANES cross-sections that have been recorded on IrO$_2$ nanoparticles of different surface areas\cite{Abbott2016}. For open-circuit conditions, the absorption edges determined for the particles with largest surface area have been found to peak at lower energies than for particles with lower surface area. Considering that the XANES cross-section of the former (latter) is expected to be more representative of surface (bulk) atoms, the shift of the white-line peak towards lower energies was interpreted as indicative of the presence of surface Ir atoms in a lower-oxidation state at low potentials\cite{Abbott2016}.

\begin{figure}
\centering
\includegraphics[width=\columnwidth]{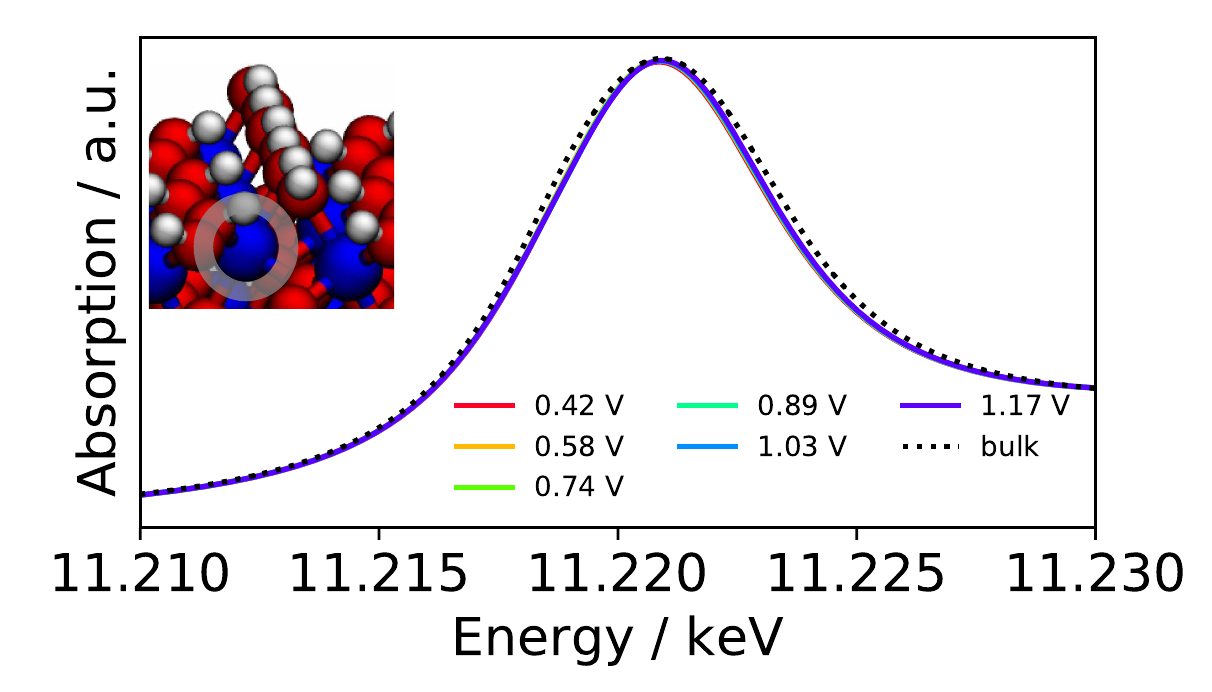}\\
\caption{Same as Figure \ref{fig:Ir-101-rec}, but the absorbing atoms are selected from the OH-covered (110) surface.}
\label{fig:Ir-101-Hall}
\end{figure}

Figure \ref{fig:Ir-101-Hall} illustrates the XANES cross-section computed for the first-layer Ir atom in the fully-OH-covered phase, which becomes the most stable (101) interface for potentials larger than $\sim0.5$ V. Similarly to what observed at the oxygen K-edge,  (see Figure \ref{fig:101-Hall}), also the iridium L$_3$-edge spectra that we have computed for the first-layer atoms of this interface very much resemble the bulk-computed cross-sections. The bulk white-line peak is only slightly broader than the surface peak, with the positions and the intensities of the two peaks matching almost exactly. Similarly, the cross-sections determined for the first-layer Ir atoms in the the OH-covered (110) surface (reported in Figure S7 in the ESI\dag) largely overlap with the bulk IrO$_2$ spectrum.

\begin{figure}
\centering
\includegraphics[width=\columnwidth]{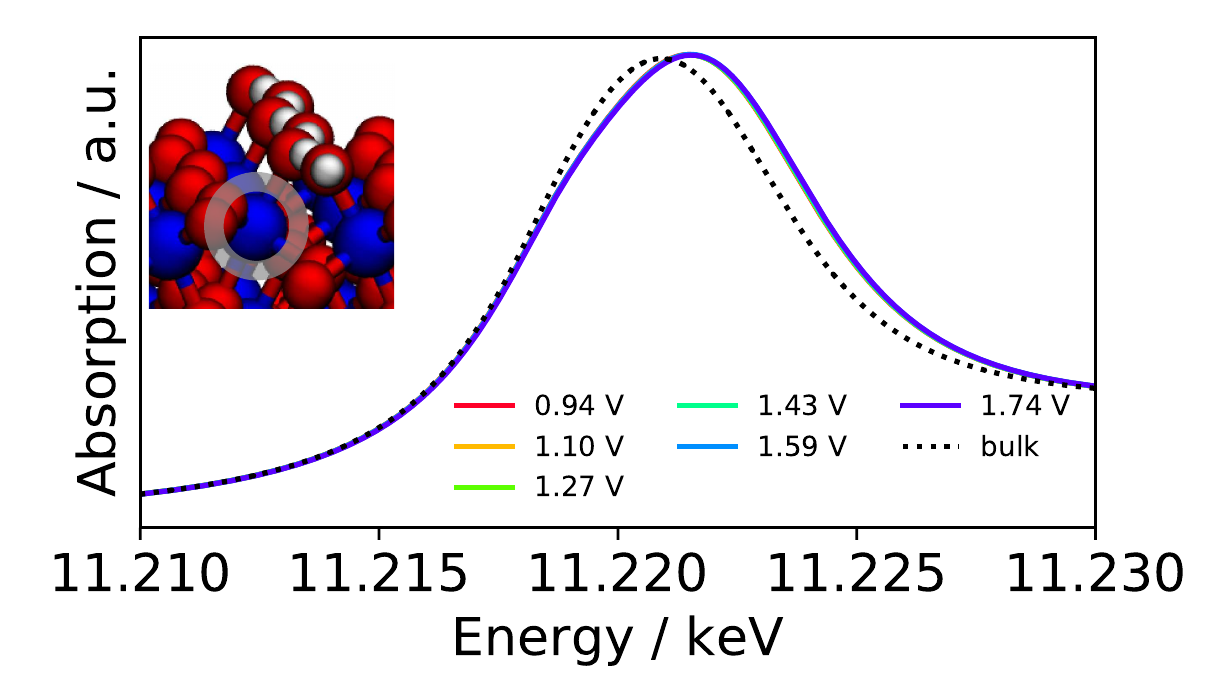}\\
\caption{Same as Figure \ref{fig:Ir-101-rec}, but the absorbing atoms are selected from a (101) termination with OH groups at the CU sites.}
\label{fig:Ir-101-H1}
\end{figure}
\begin{figure}
\centering
\includegraphics[width=\columnwidth]{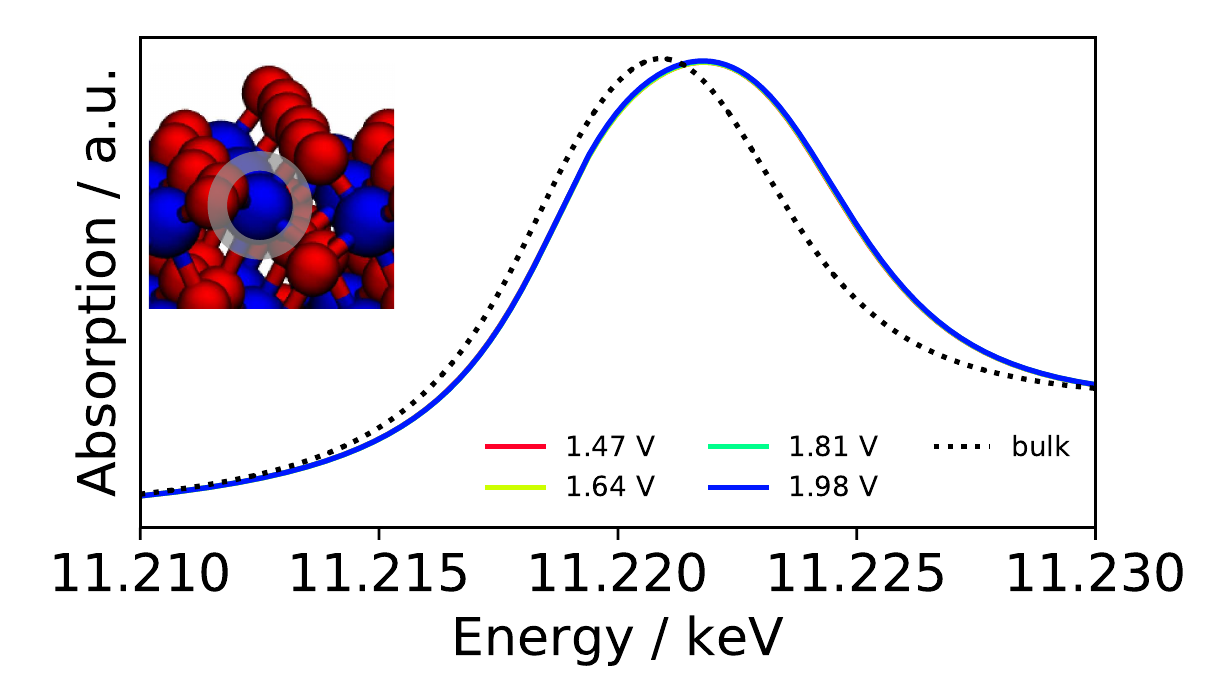}\\
\caption{Same as Figure \ref{fig:Ir-101-rec}, but the absorbing atoms are selected from an oxygen-rich (101) termination.}
\label{fig:Ir-101-OX}
\end{figure}

The absorption edge progressively shifts towards larger energies if we consider the subsequent deprotonation of the hydroxy groups at the surface. Figure \ref{fig:Ir-101-H1} and Figure \ref{fig:Ir-101-OX} show the cross-sections computed for the first-layer atoms in the (101) surface where only the bridge oxygen is deprotonated and where both bridge- and CU-site oxygens are deprotonated, respectively. The intensity of the peak remains essentially unchanged, but the absorption edge slightly broadens and shift towards higher energies. 

The XANES cross-sections computed for the (110) surface follow the same trend as for the (101) termination, with the peak  progressively shifting towards higher energies for the subsequent deprotonation of the surface hydroxyl groups (see Figure S8 and S9 in the ESI\dag). In particular, the iridium atom that is coordinated to the CU-site oxygen is the one for which we observe the strongest deviation from the bulk-computed absorption spectra. The same pattern is also observed for the (111) termination, for which the cross-section computed for the Ir atom coordinated to the CU-site oxygen is the one that is mostly shifted from the bulk spectrum (also shown in the ESI\dag, see Figure S10). Note that the CU-oxygen atoms also presented the lowest-energy absorption edges among the computed oxygen K-edge spectra. These findings suggest that the electron depletion that characterize the CU-site O atom is also shared by the underlying Ir atom, which is in a somewhat higher oxidation state.  

Note that for the reported XANES cross-sections we generally observe no potential dependence. Overall, the voltage effects that we compute are much smaller than at the O K-edge. The small potential-related final-state shifts for Ir, in fact, are to large extend covered by the convolution with a Lorentzian function with a considerably larger broadening parameter. A very minor dependence is only observed in the intensity of the white-line peak of the iridium atom underlying the CU-site oxygen in the (111) termination (see Figure S10 in the ESI\dag). 

Our results thus suggest that the hydroxylated surface that is the most stable interface at 1 V is gradually oxidized when increasing the potentials towards OER-relevant conditions, such that above $\sim1.5$ V all surface OH groups are deprotonated.   The process is predicted to be accompanied by a gradual shift of the Ir absorption edge towards higher energies. These results qualitatively agree with the findings from \emph{operando} XANES investigations on IrO$_2$ nanoparticles at the iridium L$_3$-edge\cite{Abbott2016}. For the largest-surface-area sample considered, for which the absorption spectra are ideally most representative of interface atoms, the absorption edge has been found to shift towards larger energies from 1 V to 1.44 V\cite{Abbott2016}. The oxidation of surface hydroxo species to oxide has been suggested to account for the difference in the spectra, which is consistent with the results of our simulations. Our calculations, however, cannot reproduce the experimentally-observed peak-intensity decrease that follows the absorption-edge shift\cite{Abbott2016}. A possible source of inaccuracy that should be explored in future work is the neglect of SOC in the final-state calculations for the Ir XANES cross-sections.  

\section{Conclusions\label{sec:conclusions}}

Summarizing, we have performed a theoretical investigation of the electrochemical stability of various IrO$_2$ interfaces and predicted the XANES `fingerprints' for selected terminations. In order to account for the effect of the potential, we have made use of a grand-canonical approach that allows to decouple voltage and pH effects, using a continuum description of the electrolyte solution to mimic the electrochemical environment and thus introducing the capability of first-principles \emph{operando} XANES. This strategy has enabled the simulation of XANES cross-sections under realistic conditions of applied potential, as suitable to simulate recent \emph{operando} XAS investigations on iridium-oxide-based OER catalysts.
 
In agreement with previous theoretical studies\cite{Ping2017, Pfeifer2016a, Pfeifer2017, Matz2017, Opalka2019}, results of our interface-stability analysis suggests that the (110) termination is among the most stable IrO$_2$ interfaces at moderate-potential conditions, while the (111) termination has the lowest surface energy at large applied potentials. However, our calculations suggest a reconstructed Ir-rich (101) termination to be the minimum-energy interface under open-circuit conditions. The XANES cross-sections computed at the oxygen K-edge are consistent with corresponding \emph{operando} studies\cite{Pfeifer2016a, Pfeifer2016b, Pfeifer2017, Saveleva2018, Frevel2019}. Our data support the interpretation of the appearance of pre-edge features at 529 eV (at $\sim$1 V) and 528 eV (at $\sim$1.4 V) with the formation of electron-deficient oxygen atoms at bridge- and CU-sites, respectively. We have also performed XANES simulations at the Ir L$_3$-edge. The process that leads fully-hydroxilated (110) and (101) surfaces to their oxidized forms is predicted to lead to Ir absorption cross-sections that progressively shift towards larger energies, also in agreement with experimental findings\cite{Abbott2016}.  

Overall, explicit charge-related effects on the XANES simulations have been found to be modest in the spectra simulated for the O K-edge and very weak or negligible for the Ir L$_3$-edge. The largest effects have been observed for the interfaces where voltage-induced structural changes take place, as e.g. in the presence of hydrogen-bonding networks. Experimentally-observed trends could be thus explained in terms of changes in the lowest-energy interface configuration, as predicted to be significantly larger.

\section*{Electronic Supplementary Information}
Electronic Supplementary Information (ESI) available: convergence tests for the absolute and relative core-electron binding energies, validation of the use of the scalar-relativistic approximation for the XANES initial state, comparison of the bulk-truncated and reconstructed Ir-rich termination considered, O K-edge and Ir L$_3$-edge spectra for additional surface terminations.

\begin{acknowledgements}
The authors acknowledge Prof. Matteo Calandra, Dr. Oliviero Andreussi, and Dr. Nicolas H\"ormann for very helpful discussions. This project has received funding from the European Union's Horizon 2020 research and innovation programme under grant agreements No. 798532. N. M. acknowledges support from the MARVEL National Centre of Competence in Research of the Swiss National Science Foundation. This work was supported by a grant from the Swiss National Supercomputing Centre (CSCS) under project ID s836.
\end{acknowledgements}

\bibliography{bibliography} 

\end{document}


\raggedbottom

\title{Supplemental Information: \emph{Operando} XANES from first-principles and its application to iridium oxide} 

\author{Francesco Nattino}
\email{francesco.nattino@epfl.ch}
\affiliation{Theory and Simulations of Materials (THEOS) and National Centre for Computational Design 
and Discovery of Novel Materials (MARVEL), \'{E}cole Polytechnique F\'{e}d\'{e}rale de Lausanne, 
CH-1015 Lausanne, Switzerland.}

\author{Nicola Marzari}
\affiliation{Theory and Simulations of Materials (THEOS) and National Centre for Computational Design 
and Discovery of Novel Materials (MARVEL), \'{E}cole Polytechnique F\'{e}d\'{e}rale de Lausanne, 
CH-1015 Lausanne, Switzerland.}

\date{\today} 

\maketitle

\newpage

\begin{figure}
\begin{centering}
\includegraphics[width=0.50\columnwidth]{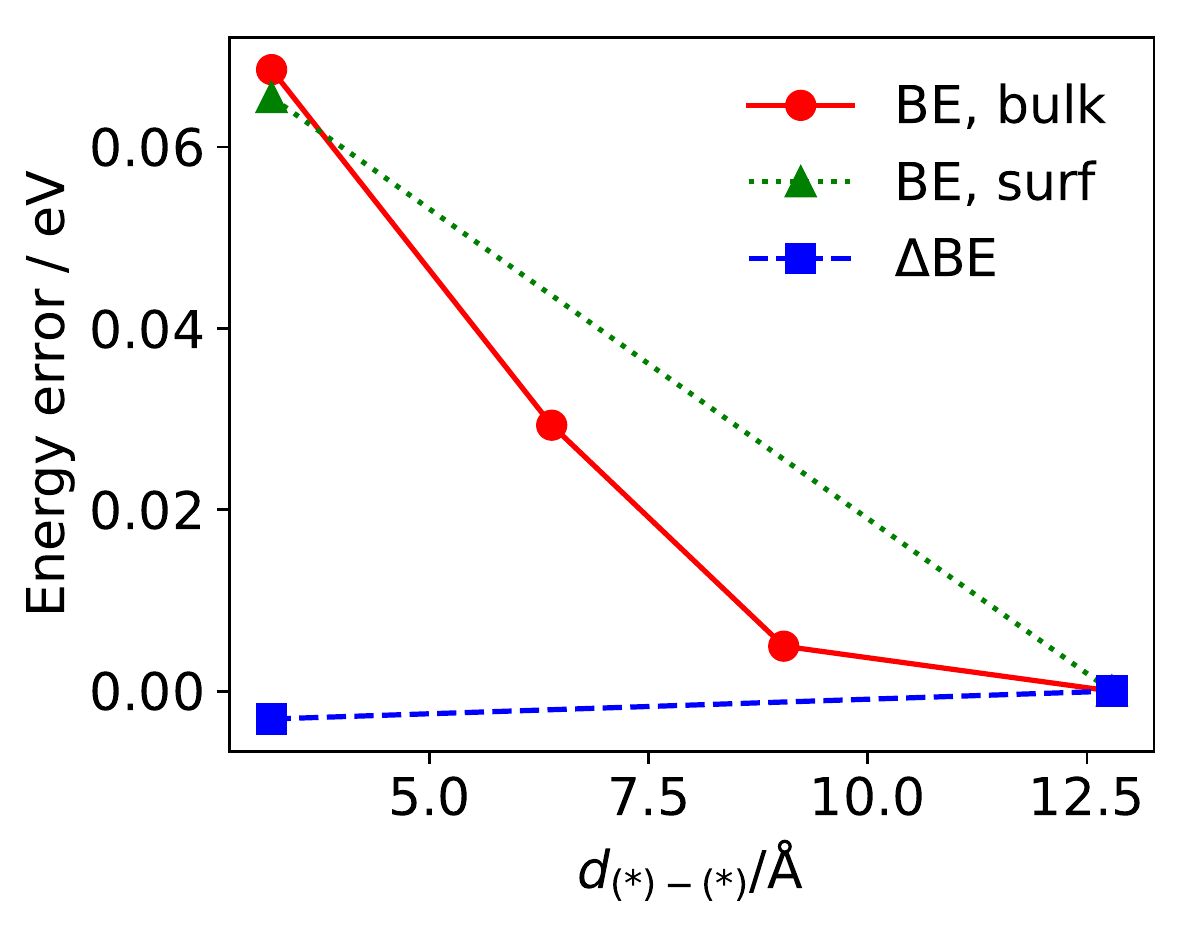}\\
\end{centering}
\caption{The convergence of the O $1s$ core-electron binding energies (BE's) is investigated as a function of the minimum distance between periodic replicas of the core-excited atom. Absolute binding energies for bulk (BE$_{\mathrm{bulk}}$, red) and surface (BE$_{\mathrm{surf}}$, green) atoms are considered, as well as the relative binding energy: $\Delta$BE = BE$_{\mathrm{surfa}}$ - BE$_{\mathrm{bulk}}$.  A (110) slab has been employed for the convergence test.  }
\label{fig:convergence-BE-OK-edge}
\end{figure}
\begin{figure}
\begin{centering}
\includegraphics[width=0.50\columnwidth]{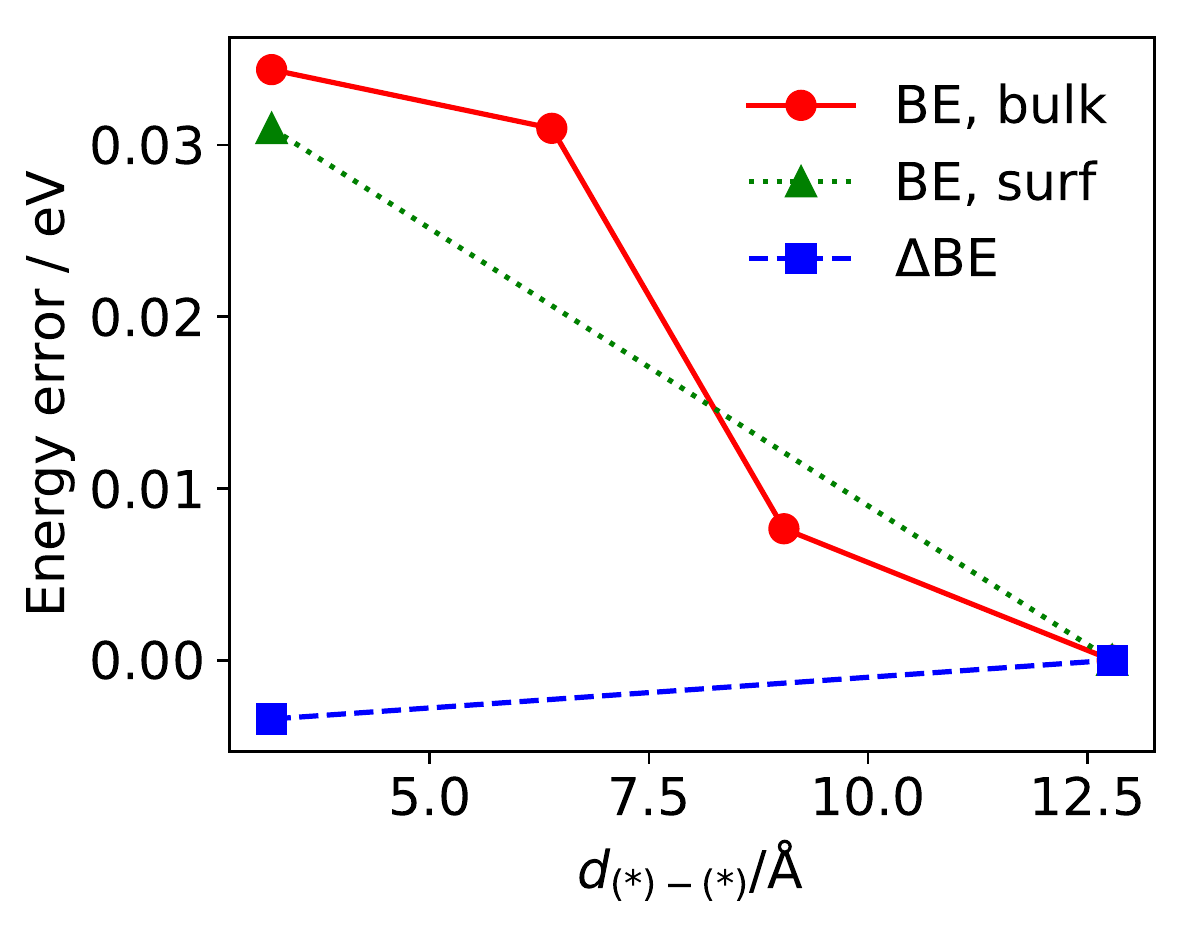}\\
\end{centering}
\caption{Same as Figure \ref{fig:convergence-BE-OK-edge}, but the Ir $2p$ core-electron BE's are considered instead.}
\label{fig:convergence-BE-IrL3-edge}
\end{figure}
\begin{figure}
\begin{centering}
\includegraphics[width=0.50\columnwidth]{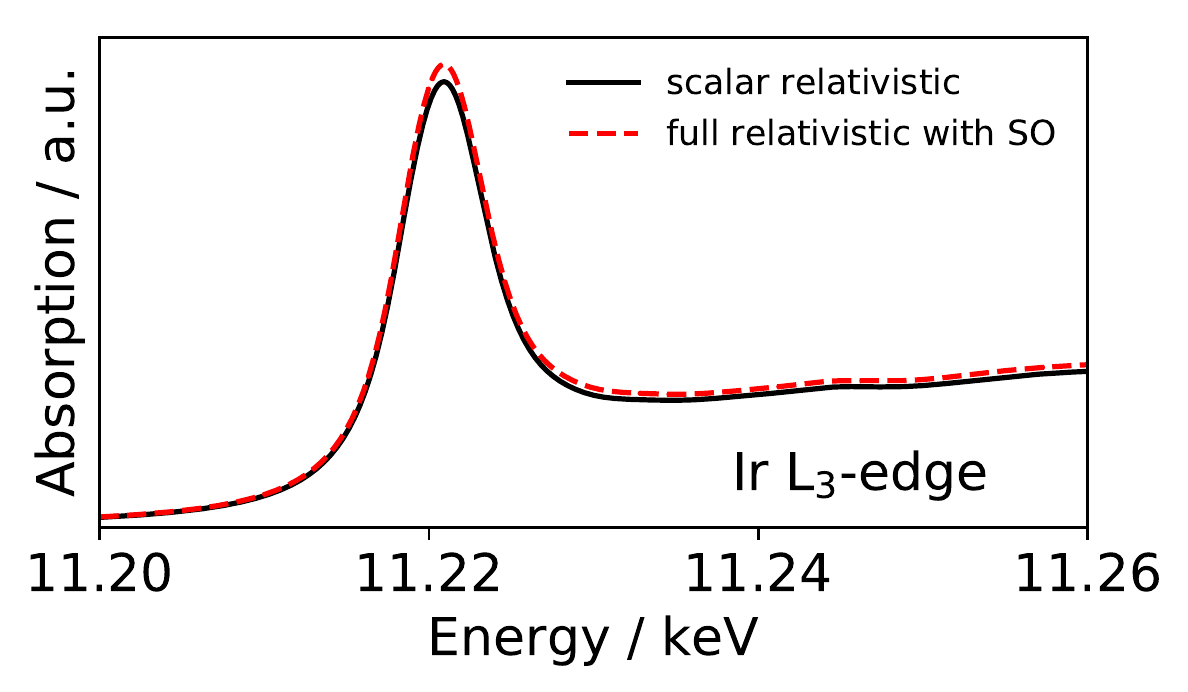}\\
\end{centering}
\caption{The Ir L$_3$-edge absorption cross-section computed using as initial state the $2p$ reconstructed all-electron wave function determined using scalar relativistic calculations (solid black) is compared to the cross-section computed using the $2p_{\frac{3}{2}}$ core state from full relativistic calculations that accounts for spin-orbit (SO) coupling (dashed red). Bulk IrO$_2$ has been employed for the test.}
\label{fig:sr-vs-fr}
\end{figure}
\begin{figure}
\begin{centering}
\includegraphics[width=0.50\columnwidth]{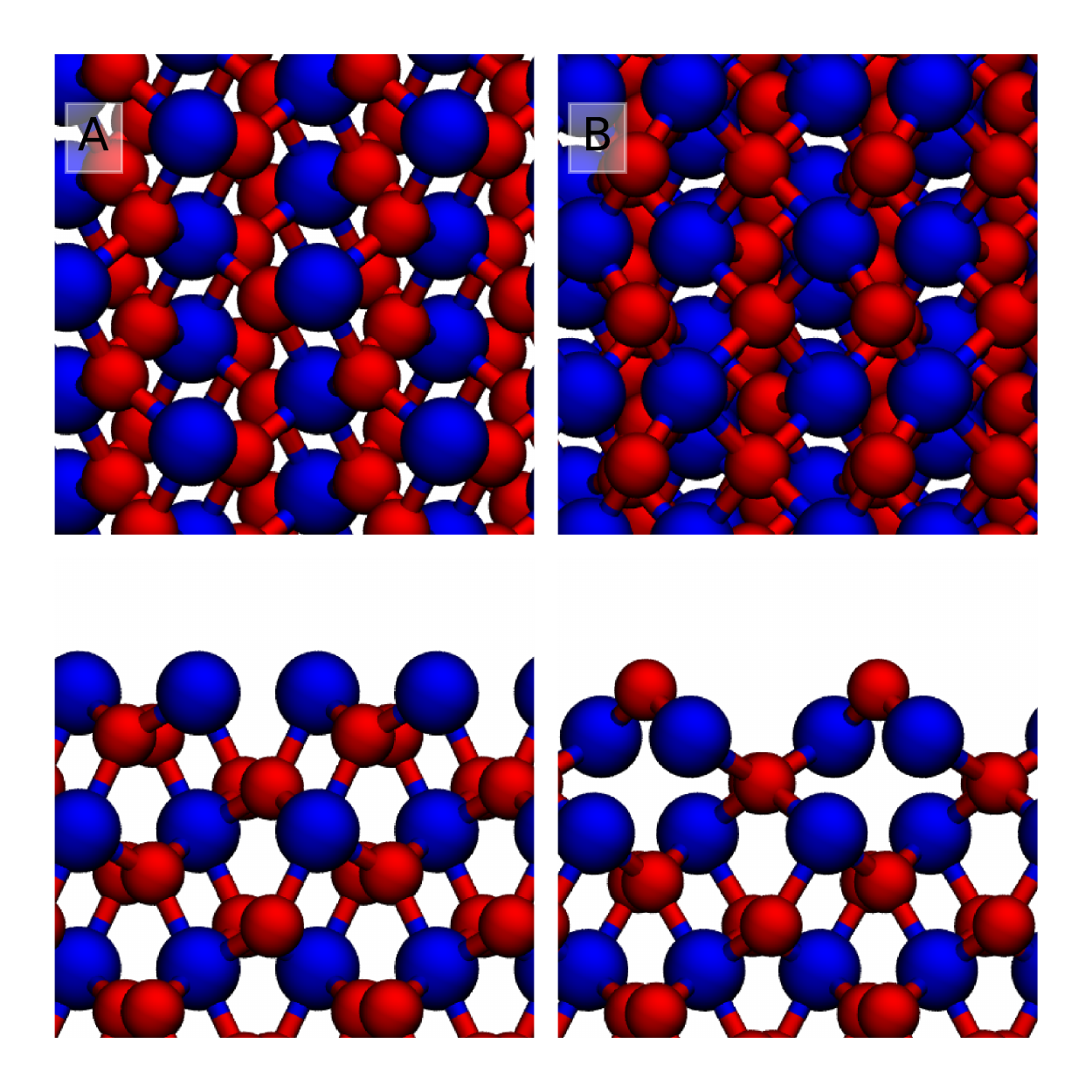}\\
\end{centering}
\caption{Top and side views of the considered Ir-rich (101) termination. Blue and red balls represent iridium and oxygen atoms, respectively. The two panels on the left (A) illustrate the bulk-truncated structure, while the reconstructed termination is sketched in the right panels (B).}
\label{fig:101-reconstruction}
\end{figure}
\begin{figure}
\begin{centering}
\includegraphics[width=0.50\columnwidth]{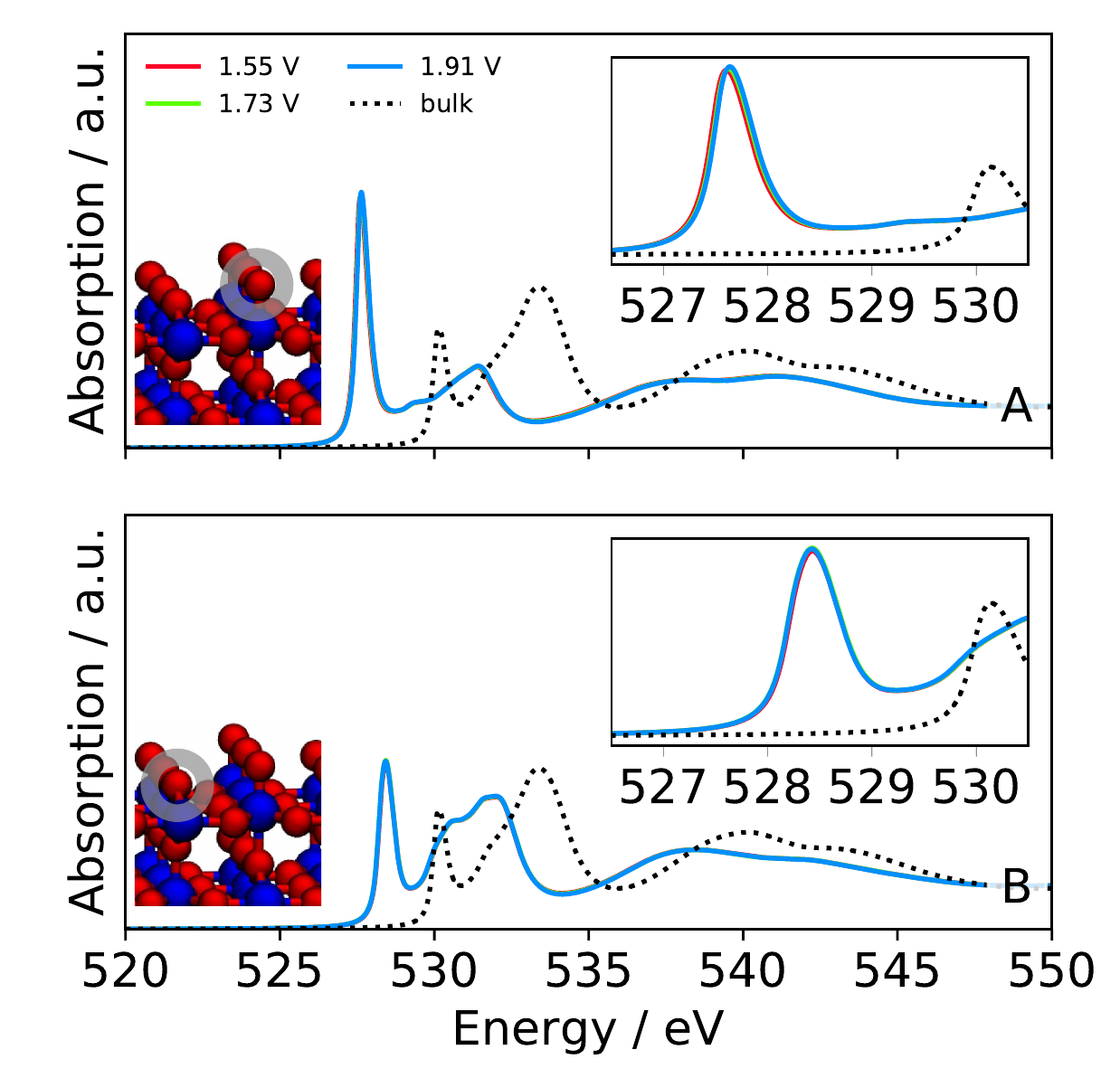}\\
\end{centering}
\caption{Oxygen K-edge XANES cross sections computed for the fully-oxidized (110) surface. A sketch of the termination is presented in the inset on the left, with blue and red balls indicating iridium and oxygen atoms, respectively. The absorbing atoms are highlighted with a grey circle. The insets on the right include a magnification of the white-line peak region. The various colors identify different potential conditions (curves are to a large extend superimposed).}
\label{fig:110-OX}
\end{figure}
\begin{figure}
\begin{centering}
\includegraphics[width=0.50\columnwidth]{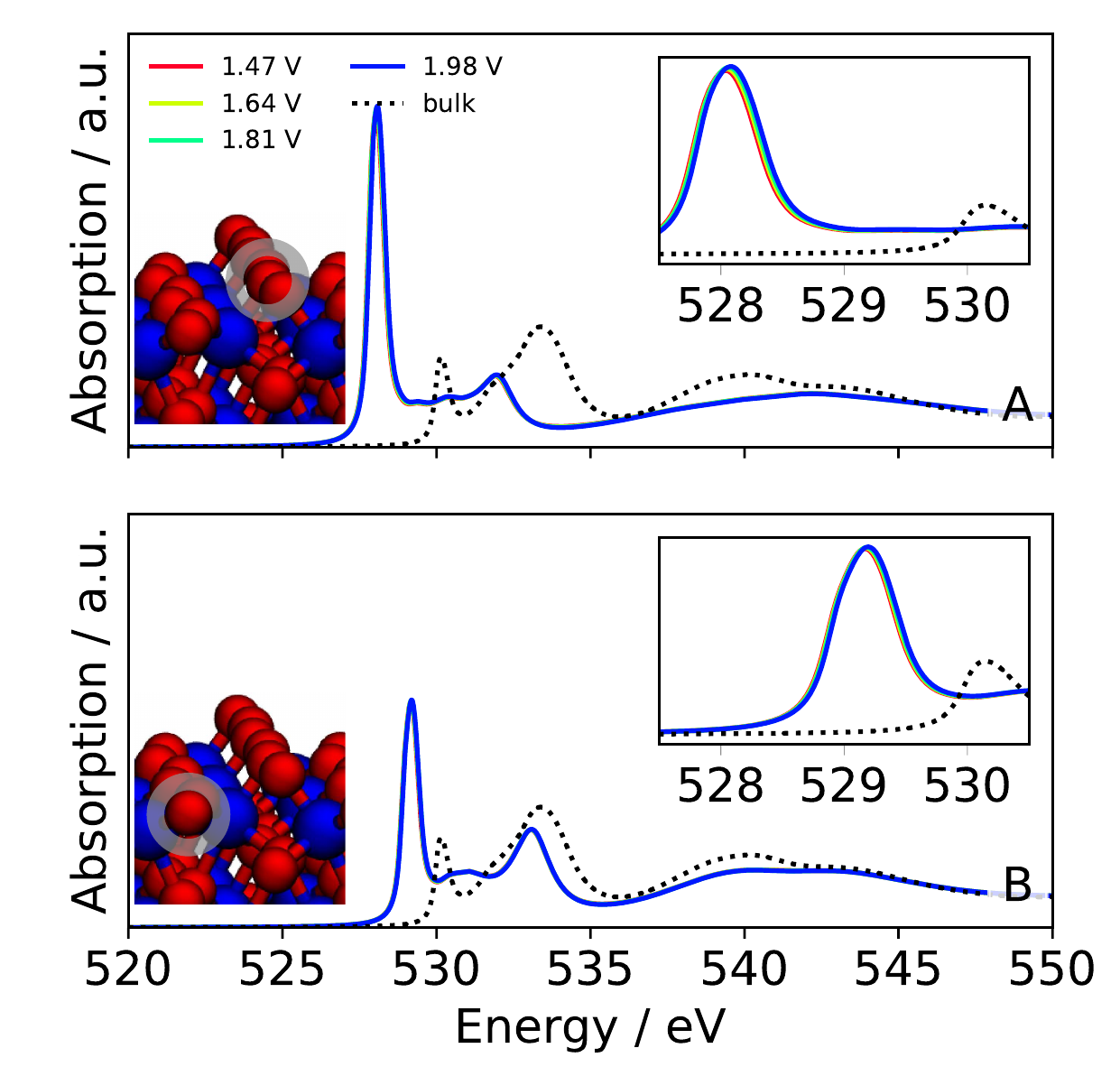}\\
\end{centering}
\caption{Same as Figure \ref{fig:110-OX}, but the absorbing atoms are selected from the fully-oxidized (101) surface.}
\label{fig:101-OX}
\end{figure}
\begin{figure}
\begin{centering}
\includegraphics[width=0.50\columnwidth]{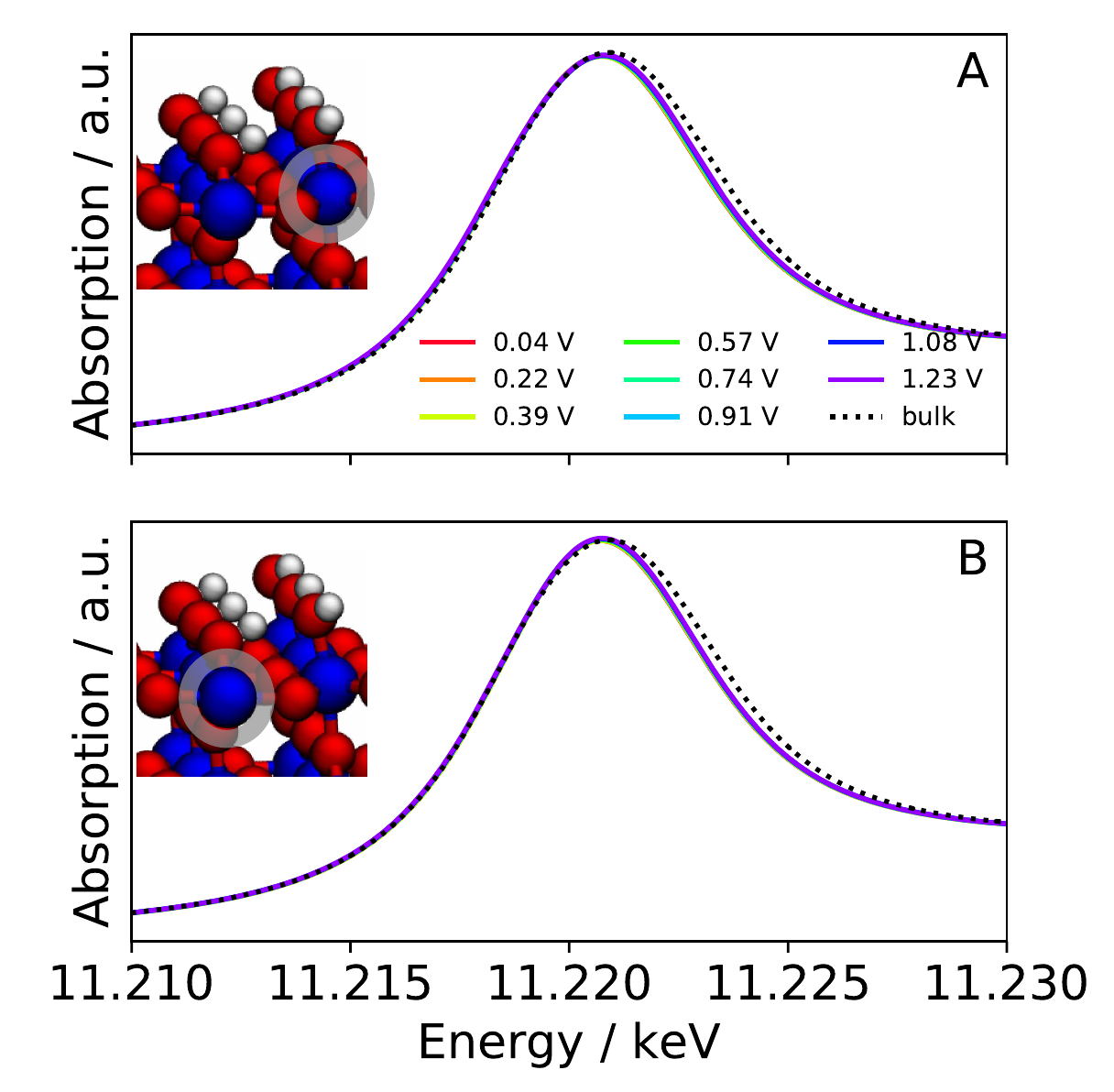}\\
\end{centering}
\caption{Iridium L$_3$-edge XANES cross sections computed for the indicated absorbing atoms of the OH-covered (110) surface. The various colors identify different potential conditions (curves are to a large extend superimposed).}
\label{fig:Ir-110-Hall}
\end{figure}
\begin{figure}
\begin{centering}
\includegraphics[width=0.50\columnwidth]{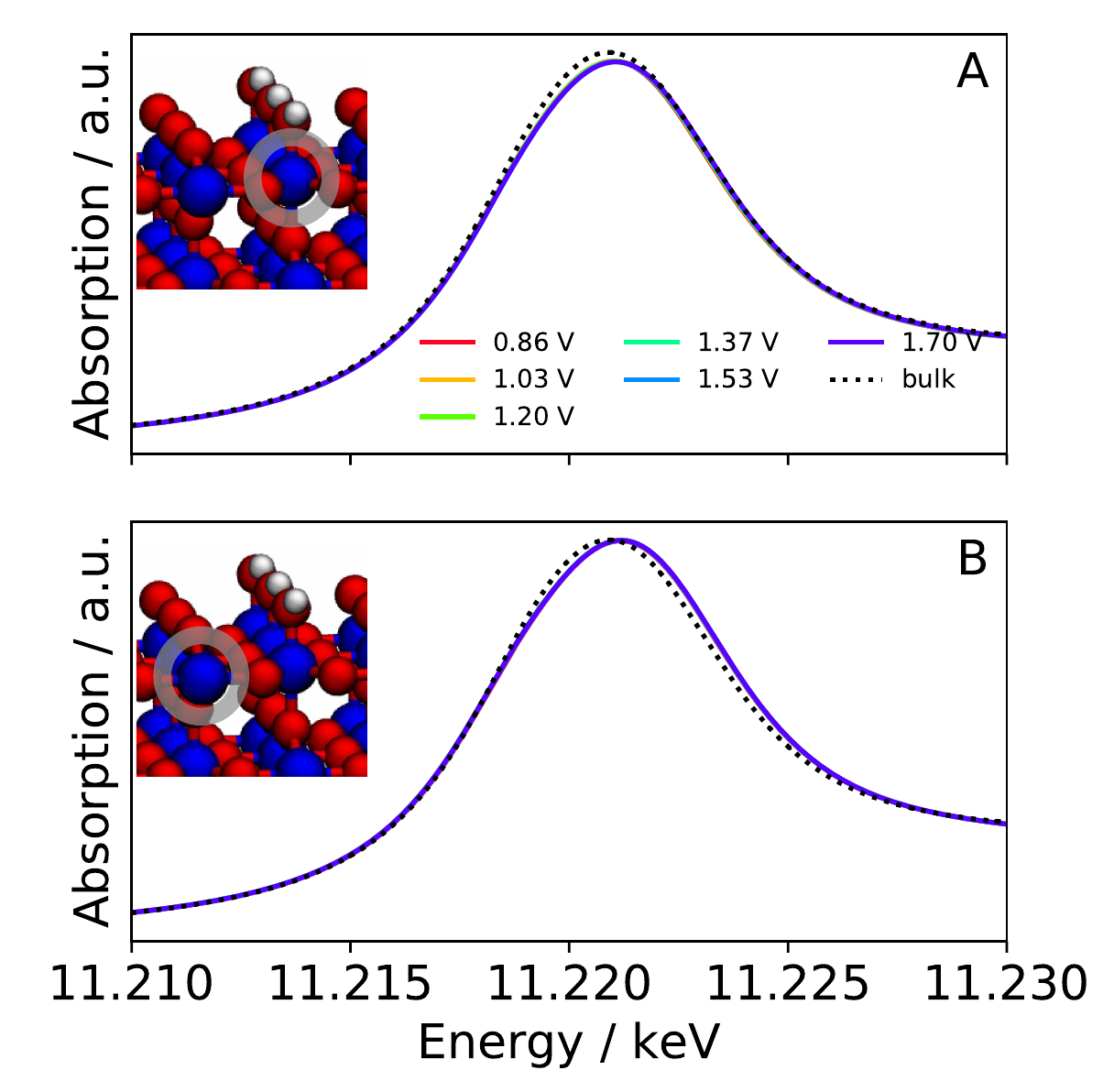}\\
\end{centering}
\caption{Same as Figure \ref{fig:Ir-110-Hall}, but the absorbing atoms are selected from a (110) termination with OH groups at the CU sites.}
\label{fig:Ir-110-H1}
\end{figure}
\begin{figure}
\begin{centering}
\includegraphics[width=0.50\columnwidth]{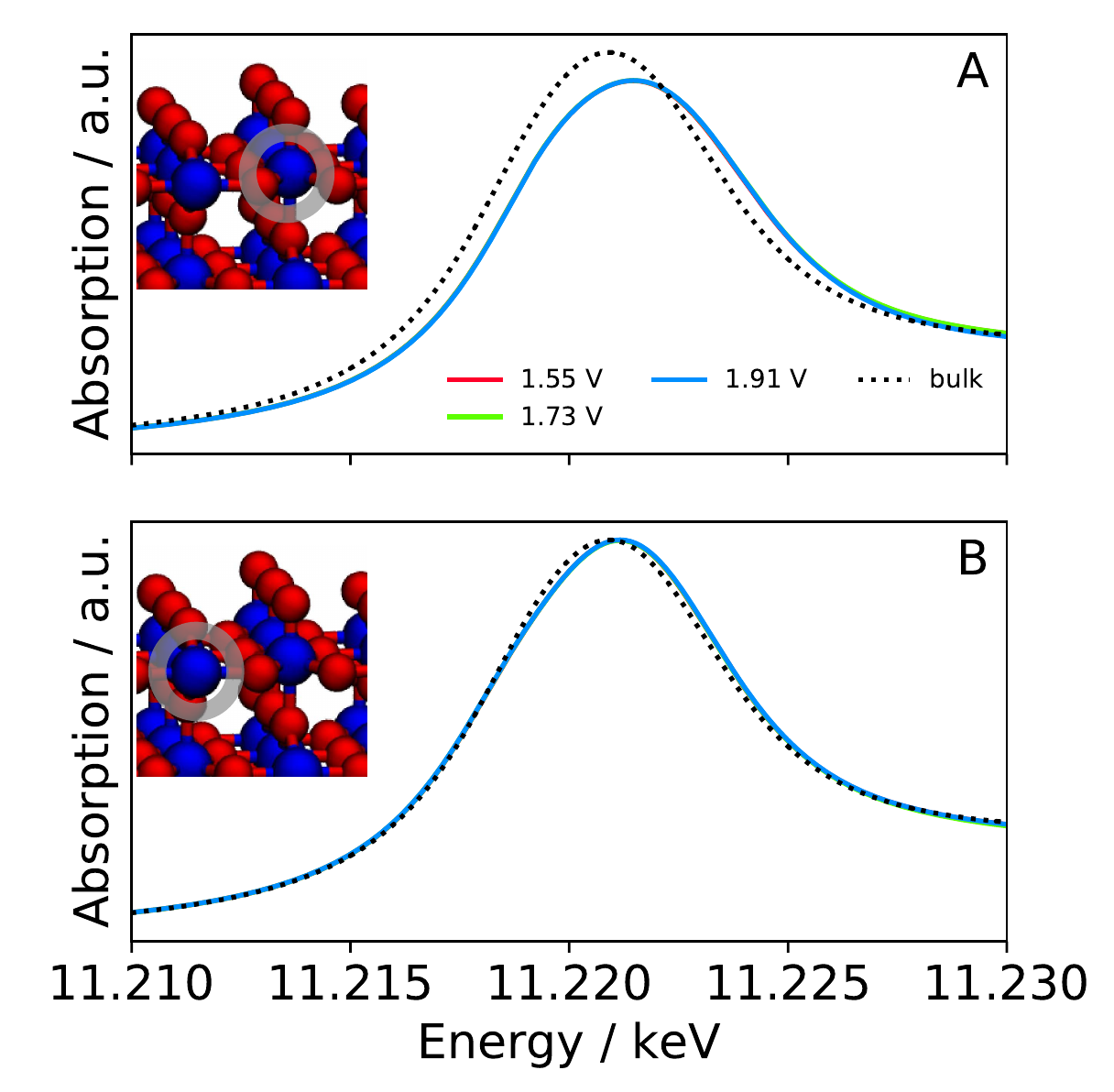}\\
\end{centering}
\caption{Same as Figure \ref{fig:Ir-110-Hall}, but the absorbing atoms are selected from the fully-oxidized (110) surface.}
\label{fig:Ir-110-OX}
\end{figure}
\begin{figure}
\begin{centering}
\includegraphics[width=0.50\columnwidth]{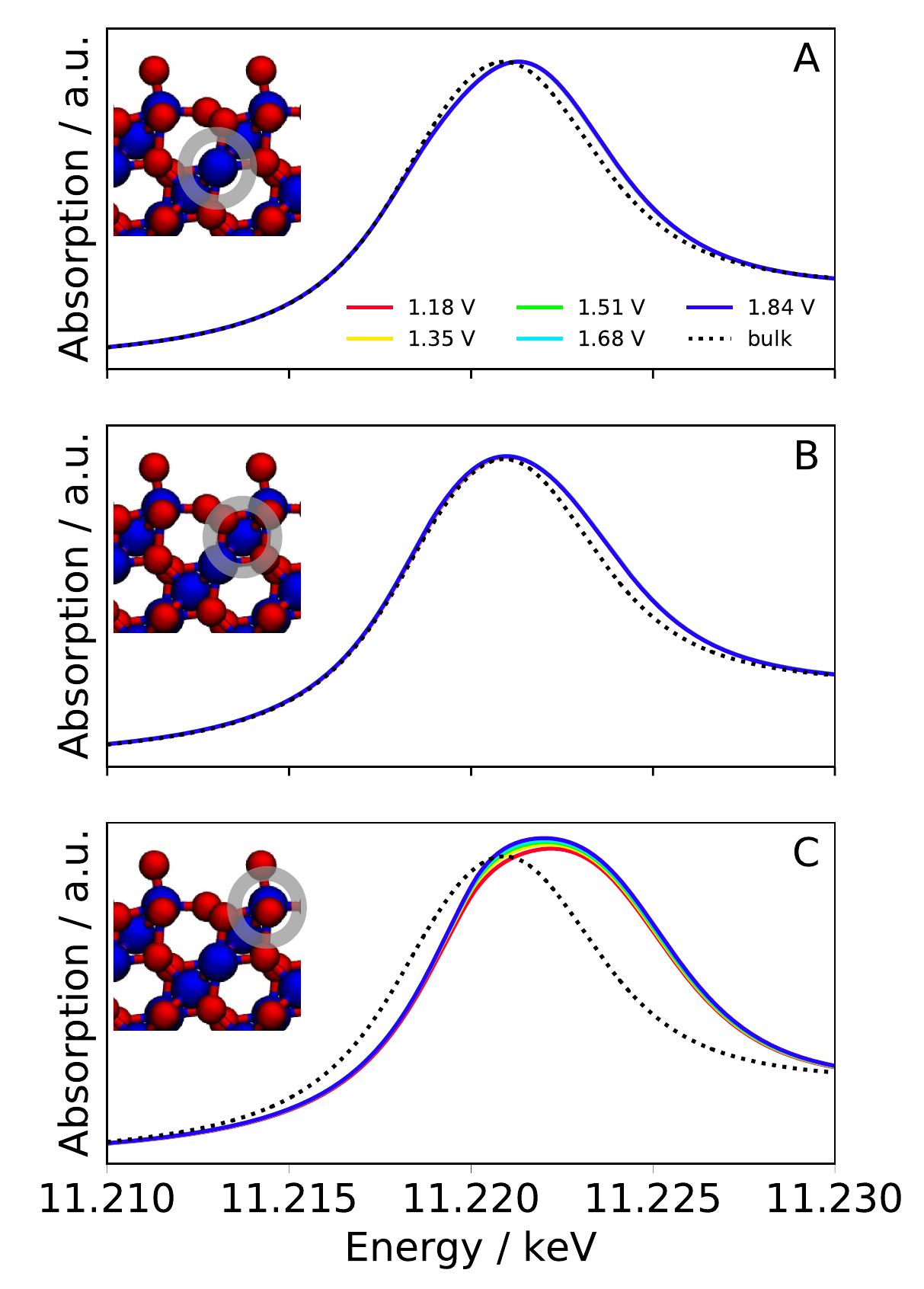}\\
\end{centering}
\caption{Same as Figure \ref{fig:Ir-110-Hall}, but the absorbing atoms are selected from a (111) termination with an oxygen atom adsorbed at the CU site.}
\label{fig:Ir-111-OX}
\end{figure}
 

